\documentclass[a4paper, twocolumn, reqno]{amsart}
\usepackage[letterpaper, portrait, margin=1in]{geometry}
\usepackage[utf8]{inputenc}
\usepackage{graphicx, caption}
\usepackage{tabularx}
\usepackage{amsfonts}
\usepackage{amsmath,bm}
\usepackage{cite}
\usepackage{comment}
\usepackage{quantikz}
\usepackage{pdflscape}
\usepackage{amssymb}
\usepackage{amsthm}
\usepackage{color}
\usepackage{mathtools}
\usepackage{braket}
\usepackage{enumitem}
\usepackage{multirow}
\usepackage{hyperref}
\usepackage{float}

\usepackage{subfig}
\usepackage{amsaddr}
\usepackage{tikz}
\usepackage{epstopdf}
\usepackage{balance}
\usepackage{algorithm}
\usepackage{algpseudocode}
\usepackage{xurl}
\usepackage[T1]{fontenc}

\title{Noise-Aware Quantum Amplitude Estimation} 
\author{Steven Herbert$^{1,2}$, Ifan Williams$^{1}$, Roland Guichard$^{1}$, Darren Ng$^{1}$}
\address{$^1$ Quantinuum, Terrington House, 13-15 Hills Rd, Cambridge, CB2 1NL,  UK  \\ $^2$ Department of Computer Science and Technology, University of Cambridge, CB3 0FD, UK}

\begin{document}
\sloppy

\onecolumn

\begin{abstract}
\noindent In this article, based on some simple and reasonable assumptions, we derive a Gaussian noise model for quantum amplitude estimation. We provide results from quantum amplitude estimation run on various IBM superconducting quantum computers and on Quantinuum's H1 trapped-ion quantum computer to show that the proposed model is a good fit for real-world experimental data. We also show that the proposed Gaussian noise model can be easily composed with other noise models in order to capture effects that are not well described by Gaussian noise. We give a generalized procedure for how to embed this noise model into any quantum-phase-estimation-free quantum amplitude estimation algorithm, such that the amplitude estimation is ``noise aware.'' We then provide experimental results from running an implementation of noise-aware quantum amplitude estimation using data from an IBM superconducting quantum computer, demonstrating that the addition of ``noise awareness'' serves as an effective means of quantum error mitigation.
\end{abstract}

\twocolumn[{%
 \centering
 \maketitle
 
 \vspace{-1cm}
}]

\section{Introduction}

\noindent {T}{he} term noisy intermediate-scale quantum computer, coined by Preskill \cite{Preskill_2018}, has become the industry standard for early quantum computers, and the name immediately tells us two of their most important features: they are noisy, and they are too small (``intermediate-scale'') to run fault-tolerant algorithms. Thus, noise as well as the consequent possibility of error, is something we have to live with. 

\indent Many ingenious suggestions for quantum error mitigation [that is, using circuit design and classical pre- and postprocessing to reduce the adverse effects of noise, without actually using any qubits for quantum error correction (QEC)] have been proposed, including zero-noise extrapolation \cite{Temme_2017, Li_2017}, randomized compilation \cite{Wallman_2016}, probabilistic error correction \cite{Endo_2018}, and many more. It is also likely that some QEC (albeit falling short of the amount required for fault tolerance) will ultimately be deployed in early algorithms exhibiting genuine useful quantum advantage (see, e.g., \cite{NISQqec, NISQqec2}), and therefore, it is perhaps more appropriate to speak of resource-constrained quantum hardware, where the aim of algorithm design is to achieve such quantum advantage with minimal resource requirements.

One approach (which, in general, is complementary to the error mitigation techniques discussed earlier) is to characterize the noise as a ``noise model'' and handle the noise at the application level. This is particularly applicable for quantum sampling algorithms (see, e.g.,  \cite{MontanaroMC, brandao2019quantum, aaronson2010computational, chowdhury2016quantum}), where the effect of the noise is that the samples are from a different distribution than in the corresponding noiseless case, but in principle with an accurate noise model, the distribution actually being sampled from can still be expressed. Moreover, sampling provides the basis for many quantum optimization and estimation algorithms \cite{MontanaroMC,brandao2019quantum}, and with an accurate noise model, there may still be quantum advantage in speed of convergence even in the presence of noise (that is, if the noise is sufficiently mild, with some errors possibly having been mitigated/corrected by some of the techniques mentioned).\\
\indent In this article, we address what is probably the simplest such instance of this concept, by proposing a suitable model for the noise accumulated when Grover iterations are repeated to perform quantum amplitude amplification. In particular, we address one such algorithm, quantum amplitude estimation (QAE), and focus on propositions for QAE, which do not call quantum phase estimation (QPE) as a subroutine \cite{Suzuki_2020, grinko2020iterative, nakaji2020faster, Aaronson_2020}. Thus, in these algorithms the quantum circuits consist only of state preparation and then amplitude amplification---there is no other circuitry involved. Moreover, these proposed algorithms all share the property that only a single qubit is measured, which gives us the basis for a simple Gaussian noise model that is valid for some particular assumptions made about the noise. In addition, QAE not only provides a suitable subject for a noise model but is also an important algorithm, which underpins quantum Monte Carlo integration (QMCI) \cite{MontanaroMC, herbert2021quantum, sjhpatent, an2020quantumaccelerated} and thus forms the basis for many important anticipated applications in, for example, quantum finance \cite{Rebentrost_2018, stamatopoulos2019option, Woerner2019, bouland2020prospects, QCfinance, egger2019credit, kaneko2020quantum, chakrabarti2020threshold, rebentrost2018quantum, financeppr, alcazar2021quantum}.

This approach of translating the noise at the physical level to the resultant estimation uncertainty at the application level in QAE has begun to gain traction recently, with both Brown et al. \cite{brown2020quantum} and Tanaka et al. \cite{tanaka2020amplitude} investigating how generic noise models, such as depolarizing, amplitude damping, and phase damping noise, impact the performance of amplitude estimation without phase estimation \cite{Suzuki_2020}. The results we present in this article are complementary to, and in some ways extend, these results. In particular, to our knowledge, this is the first time that a noise model specific to QAE has been derived, even though the aforementioned papers note the importance (and difficulty) of doing so (``It is indeed a difficult challenge to model a noise effect'' \cite[Sect.~5]{tanaka2020amplitude}; ``... any lack of accuracy in one's noise characterization will translate to a lack of accuracy in amplitude estimation. Understanding better this difficulty represents a fruitful line of future research.''  \cite[Sect. IV-A]{brown2020quantum}). Moreover, a noise model is a statistical characterization of some physical process, and hence stands independently of any particular flavor of QAE, and instead provides a quantification of the measurement uncertainty that can then be used to more accurately infer the amplitude from the measured data. To this end, our main example for how to use the proposed noise model to achieve noise-aware QAE concerns a quantification of the number of additional shots that are needed---given a maximal circuit depth at which it is possible to constrain a level of intrinsic uncertainty that arises in the model formulation---to achieve the same total mean square error (MSE) as in the noiseless case. This applies across the board to all QPE-free QAE algorithms that use repeated shots of the same circuit to infer the amplitude, including some suggestions for shallow-depth QAE \cite{giurgicatiron2020low, QCWpatent}.

Looking more broadly at QPE-free QAE, Wang et al. \cite{Wang_2021} take a different approach and cast QAE as an instance of generic observable measurement. They take an adaptive Bayesian approach, showing that by tuning the circuit parameters adaptively as the parameters become known with greater confidence, in turn, the future expected information gain can be maximized. In future work, it will be interesting to see if, and to what extent, the bespoke noise model we propose here can further enhance the Bayesian approach to quantum sampling---however, this is beyond the scope of the present paper. In addition, there are instances of QAE algorithms that do not follow the structure of repeated shots of the same circuit, for example those that require single shots of many different circuits such as the recent algorithm by Akhalwaya et al. \cite{akhalwaya2023modular}, which uses linear combination of unitary (LCU) operations to prepare initial states, leading to improved statistical robustness of the final amplitude estimate. It should be possible in future to adapt the methodology for these cases--however, this is also beyond the scope of the present paper. Finally, it is also worth noting that efforts are afoot to investigate the performance of QAE on noisy quantum computers, even without application-level handling of the noise \cite{Rao:20}.

\subsection{Article organization}

The rest of this article is organized as follows: in Section~\ref{mod} we describe in detail how the Gaussian noise model is derived from some reasonable initial assumptions, demonstrating that depolarizing noise is a special case of the Gaussian noise model and also that the Gaussian noise model can be easily composed with other noise channels to form extended models. In Section~\ref{exp}, we present a series of experiments, run on actual quantum hardware, to probe the real-world applicability of our proposed noise model. Then, in Section~\ref{res}, we include and discuss the results of these experiments. In Section~\ref{qaecal}---with a view to proposing our noise-aware QAE algorithm later---we discuss a generalized procedure for collecting and analyzing additional QAE calibration data, which we will see is a key requirement for actually running noise-aware QAE. In Section~\ref{naqae}, we then give the theory and general framework for this proposed noise-aware QAE algorithm. In Section~\ref{nares}, we give some further numerical results for the noise-aware QAE algorithm. Finally, Section~\ref{conc} concludes this article.

\section{Theoretical Derivation of the Gaussian Noise Model}
\label{mod}

\noindent To derive the Gaussian noise model, it is first necessary to introduce QAE \cite{brassard2000quantum}: QAE uses a generalization of Grover's search algorithm \cite{Groversearch}, amplitude amplification, to estimate the amplitude, $a = \sin^2 \theta$, of a general $n$-qubit quantum state expressed in the form
\begin{equation}
\label{eqn10}
\ket{\psi} = \cos\theta \ket{\Psi_0} \ket{0} + \sin\theta \ket{\Psi_1} \ket{1}
\end{equation}
for some $(n - 1)$-qubit states $\ket{\Psi_0}$ and $\ket{\Psi_1}$. A circuit, $A$, which prepares $\ket{\psi}$, that is $\ket{\psi} = A \ket{0^n}$, is taken as the input to the QAE algorithm, from which it is possible to build an operator $Q = -AS_0A^{-1}S_{\chi}$ (where $S_0 = X^{\otimes n} (C_{n-1}Z) X^{\otimes n}$ and $S_{\chi} = I_{2^{n-1}} \otimes Z$ do not depend on $A$) which performs Grover iteration
\begin{equation}
\label{eqn20}
        Q^m \ket{\psi} \! = \! \cos( ( 2m+1 ) \theta ) \! \ket{\Psi_0} \! \ket{0} \! + \! \sin((2m+1)\theta) \! \ket{\Psi_1} \! \ket{1}.
\end{equation}
Thus, for $m$ applications of $Q$, the probability of measuring the state $\ket{1}$ on the last qubit is $\sin^2((2m+1) \theta)$, and the essential idea common to all QPE-free QAE algorithms is to run circuits for a variety of different values of $m$ and then to use classical postprocessing to estimate $\theta$ and hence $a$. In the presence of noise, each Grover iterate will not necessarily enact a rotation of exactly $2 \theta$, and thus we can write the actual rotation enacted by the $i^{th}$ Grover iterate as $2 \theta + \epsilon_i$, where $\epsilon_i$ is some error, which then gives
\begin{align}
\ket{\psi} \xrightarrow[ ]{Q^m} &  \cos \left( \left(2m+1 \right) \theta + \sum_{i=1}^m \epsilon_i \right) \ket{\Psi_0}\ket{0} \nonumber \\
\label{eqn30}
& + \,\,\, \sin \left( \left(2m+1 \right) \theta + \sum_{i=1}^m \epsilon_i \right) \ket{\Psi_1}\ket{1}.
\end{align}

Note that \eqref{eqn30} is a general expression representing the final state of a QAE circuit (after $m$ Grover iterations) run on a noisy machine that characterizes the effects of most sources of noise: that is, by treating each Grover iteration to be a rotation by the desired (noiseless) angle plus some additional erroneous components. However, it is worth stating that there are certain types of noise channel that one would not necessarily expect to be modeled in this manner, and this is discussed in detail in Section~\ref{ampdampsec}. It is also pertinent that, by treating the Grover iteration as a ``closed box'' in this way, \eqref{eqn30} is agnostic to the specific realization of the algorithm as a quantum circuit---covering even cases where QEC is deployed, for example. In order to turn this into a useful noise model, it is necessary to introduce some assumptions, namely that the various $\epsilon_i$ are independent and identically distributed (i.i.d.), and importantly arise from noise effects that are independent of the quantum state itself. In the case of applications of $Q^m$ in quantum amplitude amplification, the fact that the entire evolution of the state occurs in the Bloch sphere spanning $\ket{\Psi_0} \ket{0}$ and $\ket{\Psi_1}\ket{1}$ means that this independence is easy to visualize: the distribution of $\epsilon_i$ does not depend on the point on the Bloch sphere that represents the state. Then, if $m$ is sufficiently large, we can invoke the central limit theorem (CLT) to approximate the sum of the random variables $\epsilon_i$ as a single random variable drawn from a Gaussian distribution. In this case, the state can be expressed \begin{align}
Q^m \ket{\psi} =  &  \cos ((2m+1)\theta + \theta_\epsilon) \ket{\Psi_0} \ket{0} \nonumber \\
\label{eqn40}
& \,\,\, + \sin ((2m+1)\theta + \theta_\epsilon) \ket{\Psi_1} \ket{1}
\end{align}
such that $\theta_\epsilon \sim \mathcal{N}(k_\mu m, k_ \sigma m)$, where $\mathcal{N}(\text{\small mean}, \text{\small variance})$ is the normal distribution and $k_\mu$ and $k_ \sigma$ are constants.
 
\indent This noise model applies because we are only interested in the error on a single qubit, and this allows us to use a classically inspired approach and invoke the CLT. The result is a noise model that is somewhat different in appearance to that which may conventionally be derived for modelling noisy quantum channels. In particular, we do not express a mixed state, but rather notice that a mixed state simply captures classical uncertainty about which pure state some quantum system is in. Thus, our proposed Gaussian noise model can be thought of as always treating the quantum state as pure, and then quantifying the relevant part of our uncertainty about what that quantum state is separately.

It is worth discussing in a bit more detail the assumptions made regarding the noise model. 

First, the assumption that the various $\epsilon_i$ are i.i.d. is actually an unnecessarily strong requirement. In fact, all that is required is that a sufficiently large number of independent underlying ``factors''\footnote{When running a QAE circuit on a noisy machine, such ``factors'' can include systematic effects such as hardware and calibration inaccuracies, systematic biases in the quantum gates or operations themselves, or effects such as crosstalk, temperature-induced errors, etc.} contribute to the randomness that a Gaussian random variable well models the value in question. This assumption is identical to that ubiquitously used in wireless communications literature to model noise as additive white Gaussian noise \cite{AWGN}. 

Second, while the noise model assumes that noise effects are independent of the quantum state itself, this assumption does not generally hold for all generic noise channels. We thus might not expect the model to be able to capture such effects. However, ``extended'' noise models can be built corresponding to composite random processes, such that effects that are not expected to be captured by Gaussian noise can also be incorporated. Such a model is introduced in Section~\ref{ampdampsec}. 

Finally, the Gaussian noise model relies on the assumption that the Gaussian iterate circuit remains in place; that is, the Gaussian noise model explicitly requires that the circuit structure is a succession of Grover iterates. It is reasonable to question whether this is a realistic assumption if the circuit is passed through a compiler that returns a functionally-equivalent, shallower circuit. In this case, one may ostensibly expect the required structure to be disturbed; however, a more detailed look at the structure of QAE circuits reveals that such an approach would still leave a repeated sequence of identical (or near-identical) circuit blocks. This is because, once the number of qubits has exceeded that for which the entire unitary could be resynthesized from scratch, all that can be done is a peep-hole-type optimization. In this case, the multicontrolled $CZ$ gate that is central to the Grover iterate circuit acts as a barrier to too much commutation of other gates past these regularly recurring multicontrolled $CZ$ gates, and hence, even a rewritten circuit would have a similar overall structure. Indeed, an alternative would simply be to pass the subcircuit $Q$ to the compiler to obtain circuit optimizations wholly contained therein while maintaining a succession of strictly identical blocks.

\subsection{Depolarizing noise is a special case of the Gaussian noise model}
\label{depolsec}

Of the existing generic noise models, our proposed Gaussian noise model is most similar to the depolarizing noise model. The depolarizing noise model supposes that each layer of quantum gates has some probability of completely depolarizing the quantum state, whereas the Gaussian noise model supposes that the state gradually depolarizes in a continuous manner. It turns out that when $k_\mu = 0$ the depolarizing noise model and Gaussian noise model are identical, as we now show. 

To begin, consider the probability of the measurement outcome being zero in depolarizing noise, using the fact that the circuit depth is approximately proportional to $m$ \begin{equation}
    p(0) = p_{(coh)}^{m} \cos^2 ((2m+1)\theta) + (1-p_{(coh)}^{m}) \frac{1}{2}
\end{equation}
where $p_{(coh)}$ is some (fixed) finite probability of the state remaining coherent (not depolarizing) that scales with the number of Grover iterates.
For comparison with the Gaussian noise model, it is easiest to express the probability of measuring 1 subtracted from the probability of measuring 0 \begin{align}
p(0) \!-  \! p(1)  = & \left( p_{(coh)}^{ m} \cos^2  ((2m+1)\theta)  \! + \! (1-p_{(coh)}^{m})\frac{1}{2} \right) \nonumber \\
-& \left(p_{(coh)}^{ m} \sin^2 ((2m+1)\theta) \! + \! (1-p_{(coh)}^{ m})\frac{1}{2} \right) \nonumber \\
\label{eqn90}
= \ & p_{(coh)}^{ m}   \left( \! \cos^2  ((2m+1)\theta)  \! - \! \sin^2  ((2m+1)\theta) \!  \right).
\end{align} We now show that the same asymptotic behavior occurs in the proposed Gaussian noise model. To do so, first, we consider the probability of the measurement outcome being $0$ when the noise is Gaussian, which can be expressed as
\begin{align}
    p(0) & = \int_{-\infty}^\infty p(0 | \theta_\epsilon) p(\theta_\epsilon) \ \mathrm{d} \theta_\epsilon \nonumber \\
    \label{eqn100}
    & =  \int_{-\infty}^\infty \cos^2 ( (2m+1)\theta+ \theta_\epsilon) \frac{e^{-\frac{(\theta_\epsilon - k_\mu m)^2}{2k_ \sigma m}}}{\sqrt{2 \pi k_ \sigma m }}  \, \, \mathrm{d} \theta_\epsilon.
\end{align} We can similarly express the probability of outcome $1$
\begin{equation}
\label{eqn110}
    p(1) = \int_{-\infty}^\infty \sin^2 ( (2m+1)\theta+ \theta_\epsilon) \frac{e^{-\frac{(\theta_\epsilon - k_\mu m)^2}{2k_ \sigma m}}}{\sqrt{2 \pi k_ \sigma m }}  \, \ \mathrm{d} \theta_\epsilon.
\end{equation} Next we subtract (\ref{eqn110}) from (\ref{eqn100}) \begin{align}
    \! p(0) \! & - \! p(1) \! \nonumber \\ 
    &=  \int_{-\infty}^\infty \cos^2 ( (2m+1)\theta+ \theta_\epsilon) \frac{e^{-\frac{(\theta_\epsilon - k_\mu m)^2}{2k_ \sigma m}}}{\sqrt{2 \pi k_ \sigma m }}   \nonumber \\
    &\,\,\,  \,\,\,  \,\,\,  \,\,\, \,\,\,  \,\,\,  \,\,\,  \,\,\,- \sin^2 ( (2m+1)\theta+ \theta_\epsilon) \frac{e^{-\frac{(\theta_\epsilon - k_\mu m)^2}{2k_ \sigma m}} }{\sqrt{2 \pi k_ \sigma m }}  \, \, \mathrm{d} \theta_\epsilon \nonumber \\
    %& = \frac{1}{\sqrt{2 \pi k_2 m }} \int_{-\infty}^\infty \left( \cos^2 ( (2m+1)\theta+ \theta_\epsilon) - \sin^2 ( (2m+1)\theta+ \theta_\epsilon) \right) e^{-\frac{(\theta_\epsilon - k_\mu m)^2}{2k_2m}}  \, \, \mathrm{d} \theta_\epsilon \nonumber \\
     & =   \frac{1}{\sqrt{2 \pi k_ \sigma  m }}  \nonumber \\
    &  \,\,\,  \,\,\,  \,\,\,  \,\,\,  \int_{-\infty}^\infty \! \cos\left( 2 \left( (2m+1)\theta+ \theta_\epsilon \right) \right)   e^{-\frac{(\theta_\epsilon - k_\mu m)^2}{2k_ \sigma m}}  \, \, \mathrm{d} \theta_\epsilon \nonumber \\
     & =  \frac{1}{\sqrt{2 \pi k_ \sigma m}}  \mathcal{R} \! \left( \! e^{2i(2m+1) \theta} \int_{-\infty}^\infty e^{2i\theta_\epsilon} e^{-\frac{(\theta_\epsilon - k_\mu m)^2}{2k_ \sigma m}}\, \, \mathrm{d} \theta_\epsilon \! \right) \nonumber \\
     & =  \frac{1}{\sqrt{2 \pi k_ \sigma m}}  \mathcal{R} \! \left( \! e^{2i(2m+1) \theta+2ik_\mu m} \! \int_{-\infty}^\infty \! e^{2i\tilde{\theta}_\epsilon -\frac{\tilde{\theta}_\epsilon^2}{2k_ \sigma m}}\ \, \mathrm{d} \tilde{\theta}_\epsilon \! \right) \nonumber \\  
     & =  \frac{1}{\sqrt{2 \pi k_ \sigma m}}  \mathcal{R} \left( e^{2i(2m+1) \theta+2ik_\mu m} \sqrt{2 \pi k_ \sigma m} \ e^{-2k_ \sigma m} \right) \nonumber \\
    %& = \mathcal{R} \left( e^{2i(2m+1) \theta+2ik_1m} \, e^{-2k_2 m} \right) \nonumber \\
    & =  e^{-2k_ \sigma m} \cos (2((2m+1) \theta+k_\mu m)) \nonumber \\
    & =  e^{-2k_ \sigma m} \Big( \cos^2 ((2m+1) \theta+k_\mu m)  \nonumber \\
    \label{eqn115}
    & \,\,\,\,\,\,\,\,\,\,\,\,\,\,\,\,\,\,\,\,\,\,\,\,\,\,\,\,\,\,\,\,\,\,\,\,\,\,\,\,\,\,  - \sin^2 ((2m+1) \theta+k_\mu m) \Big). 
\end{align} where $\mathcal{R}$ denotes the real part, and the substitution $\tilde{\theta}_\epsilon = \theta_\epsilon - k_\mu m$ is used to simplify the integral.\\
\indent We can see that when $k_\mu =0$ (which is effectively implicit in the depolarizing noise model), the depolarizing and Gaussian noise models have identical measurement statistics, that is, when the parameters are matched accordingly ($\tilde{p}_{coh} \equiv e^{-2 k_\sigma}$), (\ref{eqn90}) and (\ref{eqn115}) are equivalent. Note that, as there are only two measurement outcomes, namely, 0 and 1, and probabilities sum to one, the measurement statistics are completely determined by the value of $p(0) - p(1)$.

This correspondence between the depolarizing and Gaussian noise models is important for a number of reasons. First, the fact that the Gaussian noise model we propose here corresponds to a widely used and accepted noise model (i.e., depolarizing noise) adds credence to its validity. Conversely, the fact that the Gaussian noise model has been derived from simple and justified physical assumptions gives a phenomenological (rather than ad hoc) basis to use the depolarizing noise model in appropriate circumstances.

\subsection{Composite noise model including the effects of amplitude damping}\label{ampdampsec}

The proposed Gaussian noise model assumes that the effects of noise are independent of the quantum state itself. However, there exists (at least) one widely used, and experimentally observed, noise channel that violates this assumption of independence, namely, amplitude damping.

Amplitude damping assumes that the probability amplitudes associated with quantum states tend to decay over time due to interactions with an external environment. Since qubit states of a machine correspond to physical states of some system with given physical properties, amplitude damping will affect the probability amplitudes differently depending on these properties. For instance, if qubit states are mapped to physical states of a quantum system that correspond to different energies, then the dynamics of noisy interactions with the environment from effects such as spontaneous emission will differ between the physical states. This will lead to an explicit dependence of the noise on the state of the quantum system itself.

Amplitude damping is characterized by a damping parameter $\gamma$, quantifying the rate at which the system loses coherence. Taking a very simplified example of a single-qubit system with qubit states $\ket{0}$ and $\ket{1}$ mapped to physical states such that the $\ket{1}$ state is mapped to a higher energy physical state---meaning in this simple example it is more likely to interact with the environment through effects such as spontaneous emission--then we can model the amplitude damping channel as affecting the amplitudes of the state such that each qubit state has a modified amplitude \begin{align}\label{eqn115.5}
& \tilde{p}(0) = p(0) + \gamma p(1)  \nonumber \\
& \tilde{p}(1) = (1-\gamma) p(1).
\end{align} As this random process is not independent of the state, the proposed Gaussian noise model does not capture the error introduced by amplitude damping, which manifests for a single qubit as a bias to the measurement statistics in favor of one of the qubit states. However, it is possible to build a noise model corresponding to a composite random process that incorporates Gaussian noise and other effects not captured by Gaussian noise such as amplitude damping, leading to a more comprehensive model.

To build a composite noise model that captures both Gaussian noise and amplitude damping, we first note that for the Gaussian noise model, $p(0) - p(1)$ is as given in (\ref{eqn115}) and also use the simple fact that $p(0)+p(1) = 1$ to obtain

\begin{alignat}{2}
p(0) &= \frac{1}{2} \Big[1+e^{-2k_{\sigma}m}  \Big(\cos^2 ((2m+1) \theta+k_\mu m) \nonumber \\ 
& \ \ \ \  \ \ \ \ \  \ - \sin^2 ((2m+1) \theta+k_\mu m)\Big) \Big] \nonumber \\
p(1) &= \frac{1}{2} \Big[1-e^{-2k_{\sigma}m}  \Big(\cos^2 ((2m+1) \theta+k_\mu m) \nonumber \\
& \ \ \ \  \ \ \ \ \  \ - \sin^2 ((2m+1) \theta+k_\mu m)\Big) \Big]. \nonumber \\ \label{prob_eqns}
\end{alignat} Then, under the physical scenario in which the $\ket{1}$ state is more likely to interact with the environment than the $\ket{0}$ state,\footnote{Such a choice is, however, arbitrary, and in reality, the choice will depend on the particular qubit mapping of the hardware. If the physical setup is such that the $\ket{0}$ state is more likely to interact with the environment, then the qubit labels can be simply swapped in these expressions.} given (\ref{eqn115.5}) and defining $\gamma = 1-e^{-k_{\text{AD}}m}$, we can compose the amplitude damping channel with Gaussian noise, such that the probability amplitudes of the state are then
modified according to \begin{alignat}{2}
\tilde{p}(0) &= \frac{1}{2} \Bigg[1+e^{-2k_{\sigma}m} \nonumber \Big( \cos^2 ((2m+1) \theta + k_\mu m) \\
& \ \ \ \ - \nonumber \sin^2 ((2m+1) \theta+k_\mu m) \Big) + (1-e^{-k_{\text{AD}}m}) \nonumber \\
& \ \ \ \ \ \Big[1-e^{-2k_{\sigma}m} \Big( \cos^2 ((2m+1) \theta+k_\mu m) \nonumber \\
& \ \ \ \ - \sin^2 ((2m+1) \theta+k_\mu m) \Big) \Big] \Bigg] \nonumber\\ 
\tilde{p}(1) &= \frac{1}{2} e^{-k_{\text{AD}}m}\Bigg[1-e^{-2k_{\sigma}m} \Big( \cos^2 ((2m+1) \theta+k_\mu m) \nonumber \\
& \ \ \ \ -  
\sin^2 ((2m+1) \theta+k_\mu m)\Big) \Bigg].\nonumber \\
\end{alignat}
It is important to note here that by building such a composite model, we are implicitly making the assumption that the effects of the amplitude damping channel can be modelled as if it was a pure channel applied to the final state, i.e., to the state after the pure Gaussian noise channel has already been applied, that is, rather than amplitude damping being continuously interleaved with Gaussian noise throughout the evolution. Even though this does not perfectly capture the true physical dynamics, empirical results given later in Section~\ref{res} demonstrate that such a model is appropriate.

In a similar manner, it is possible to rigorously compose the Gaussian noise model with any other independent noise channel (or several such channels), as long as one can also make the assumption that the effects of the noisy channels can be modelled as if occurring purely to the initial or final state i.e., before or after the Gaussian noise channel has been applied. One such example of additional effects not expected to be captured by Gaussian noise that can be modelled in this way is ``state-preparation and measurement'' (SPAM) errors. State-preparation errors arise from nonidealities in initializing quantum states, e.g., some qubits assumed to be initialized in the \(\ket{0}\) state may not be, and measurement errors refer to imperfections in the readout process, e.g., where a final state, such as \(\ket{1}\), is incorrectly measured as \(\ket{0}\). However, the reason that the Gaussian model composed with the amplitude damping channel is studied in particular is based on preliminary studies, which demonstrated that amplitude damping effects were clearly observed in the data, whereas the contributions from other sources were observed to generally be negligible.\footnote{However, contribution from SPAM errors in particular may have been observed for some of the data (see Section~\ref{res}).}

The ability to compose the Gaussian noise model with other channels means that arbitrary noise effects that are not well captured by Gaussian noise can be included in an extended model. This significantly increases the utility of the Gaussian noise model.

\section{Noise-Model Experimental Setup}
\label{exp}

\noindent In order to gather experimental results to assess the real-world applicability of the proposed Gaussian noise model, it was necessary to select some state-preparation circuits, $A$, and run QAE on real quantum hardware. A number of different state-preparation circuits were thus defined, and for each round of experiments, given subsets of these circuits were run (depending on the particular quantum hardware that was run and the dates that the experiments were carried out).

Generally, when designing the state-preparation circuits, we were guided by the following two principles: first, we minimized the number of qubits used in $A$ in order that we could run deeper circuits (more applications of $Q$) which thus allowed us to better observe how well the Gaussian noise model fits the experimental data; second, we chose circuits that have the property that noiselessly $\theta$ is such that on periodic numbers of Grover iterations, the measurement outcome is either $\ket{0}$ or $\ket{1}$ with certainty, which allowed us to display plots to illustrate how well the noise model fits the experimental data. Two circuits that achieved these aims are $A_1$ and $A_2$, shown in Fig.~\ref{Acircs}, which prepare the states
\begin{align}
\label{A1defeqn}
        A_1\ket{00} = & \cos(\pi/6)\ket{\Psi_0}\ket{0} + \sin(\pi/6)\ket{\Psi_1}\ket{1}\\
        A_2\ket{00} = & \cos(\pi/3)\ket{\Psi_0}\ket{0} + \sin(\pi/3)\ket{\Psi_1}\ket{1}.
\end{align}
Therefore, we can see that
\begin{align}
        Q_1^{m} A_1\ket{00} = & \cos((2m+1)\pi/6)\ket{\Psi_0}\ket{0} \nonumber \\
        & \,\,\,\,\,\,\,\,\, + \sin((2m+1)\pi/6)\ket{\Psi_1}\ket{1}\\
        Q_2^{m} A_2\ket{00} = & \cos((2m+1)\pi/3)\ket{\Psi_0}\ket{0} \nonumber \\
        & \,\,\,\,\,\,\,\,\, + \sin((2m+1)\pi/3)\ket{\Psi_1}\ket{1}
\end{align}
where $Q_1$ and $Q_2$ are the Grover iterate circuits for $A_1$ and $A_2$, respectively. We can see that $Q_1^{m} A_1 \ket{00}$ (respectively, $Q_2^{m} A_2 \ket{00}$) has the property that the measurement of the last qubit is $1$ (respectively, $0$) with certainty when $(m-1) \mod 3 = 0$ (that is for $m=1,4,7,...$). Fig.~\ref{expfig1} illustrates this for the case of $A_1$. Even though in principle QAE could be applied to a single-qubit circuit, $A$, the lack of entanglement therein would make it an unsuitable experiment, and thus, by selecting two-qubit circuits, we have used the minimum number of qubits that we can reasonably expect to lead to meaningful results.

In addition, a three-qubit circuit, $A5$, was also defined, which prepares the state
\begin{equation}
\label{A5defeqn}
A_5\ket{000} = \cos(\pi/6)\ket{\Psi_0}\ket{0} + \sin(\pi/6)\ket{\Psi_1}\ket{1}.
\end{equation} $A_5$ has the same ``periodic'' property as $A_1$, where noiselessly $Q^{m} A_5 \ket{000}$ yields measurement outcome $1$ with certainty on the third qubit when $(m-1) \mod 3 = 0$. 

\indent While these circuits achieved our aim of providing results that can readily be plotted, it was also beneficial to include a range of other circuits. To this end, we also ran QAE for $A_1$ and $A_2$ simultaneously (to introduce the possibility of crosstalk) and additionally defined the circuits $A_3$ and $A_4$, also shown in Fig.~\ref{Acircs}, which prepare the states
\begin{align}
        A_3\ket{00} = & \cos(1/2)\ket{\Psi_0}\ket{0} + \sin(1/2)\ket{\Psi_1}\ket{1}\\
        A_4\ket{00} = & \cos(1)\ket{\Psi_0}\ket{0} + \sin(1)\ket{\Psi_1}\ket{1}.
\end{align}
$A_3$ and $A_4$ are such that the final qubit does not align with the $\ket{0}$ or $\ket{1}$ axes after any number of applications of $Q$.
\begin{figure}[t!]
\captionsetup[subfloat]{position=bottom,labelformat=empty}
\centering
\subfloat[$A_1$]{\begin{quantikz}
  \qw & \gate{R_y(\frac{\pi}{3})} & \ctrl{1}& \gate{R_y(0.13\pi)} & \qw  \\
  \qw & \qw  & \gate{R_y(\pi)} & \qw & \qw \\
\end{quantikz}}
\newline
\subfloat[$A_2$]{\begin{quantikz}
  \qw & \gate{R_y(\frac{2\pi}{3})} & \ctrl{1}& \gate{R_y(0.91\pi)} & \qw  \\
  \qw & \qw  & \gate{R_y(\pi)} & \qw & \qw \\
\end{quantikz}}
\newline
\subfloat[$A_3$]{\begin{quantikz}
  \qw & \gate{R_y(1.00)} & \ctrl{1}& \gate{R_y(0.13\pi)} & \qw  \\
  \qw & \qw  & \gate{R_y(\pi)} & \qw & \qw \\
\end{quantikz}}
\newline
\subfloat[$A_4$]{
\begin{quantikz}
  \qw & \gate{R_y(2.00)} & \ctrl{1}& \gate{R_y(0.91\pi)} & \qw  \\
  \qw & \qw  & \gate{R_y(\pi)} & \qw & \qw \\
\end{quantikz}}
\newline
\subfloat[$ \ \ \ \ \ \ \  A_5$]{
\ \ \ \ \ \begin{quantikz}
  \qw & \gate{R_y(\frac{\pi}{4})} & \qw & \ctrl{2} & \qw  \\
  \qw & \gate{R_y(\frac{\pi}{4})}  & \ctrl{1} & \qw & \qw \\
  \qw & \qw  & \targ{} & \targ{} & \qw \\
\end{quantikz}}
\newline
\captionsetup{width=.9\linewidth}
	\caption{State-preparation circuits.}
	\label{Acircs}
\end{figure}
 \begin{figure}[t!] 
\centering

\begin{tikzpicture}[scale = 0.666]
\draw[line width=0.5mm, gray,-latex] (-4.5,0) -- (4.5,0);
\draw[line width=0.5mm, gray, -latex] (0,-3.5) -- (0,3.5);
\node at (5,0) {\Large $\ket{0}$};
\node at (0,4) {\Large $\ket{1}$};
\draw[line width=0.75mm, black, latex-latex] (-2.5, -1.2058697677525623) -- (2.5,1.2058697677525623);
\node at (3.3,1.2) {\Large $\ket{\psi}$};
\node at (-3.5,-1.2) {\Large $Q_1^3\ket{\psi}$};
\draw[line width=0.75mm, black, latex-latex] (0, -2.5) -- (0, 2.5);
\node at (1.25,2.5) {\Large $Q_1\ket{\psi}$};
\node at (1.25,-2.5) {\Large $Q_1^4\ket{\psi}$};
\draw[line width=0.75mm, black, -latex] (0, 0) -- (-2.5,1.2058697677525623);
\node at (-3.5,1.2) {\Large $Q_1^2\ket{\psi}$};
\draw[thick] (0.5,0) arc (0:270:0.5);
\node at (1.5, 0.3) {$\frac{\pi}{6}$};
\node at (0.5, 1) {$\frac{\pi}{3}$};
\node at (-0.5, 1) {$\frac{\pi}{3}$};
\node at (-1.5, 0) {$\frac{\pi}{3}$};
\node at (-0.5, -1) {$\frac{\pi}{3}$};
\end{tikzpicture}

\captionsetup{width=0.9\linewidth}
\caption{Illustration of the action of $Q_1^m$ on $\ket{\psi} = A_1 \ket{00}$. Here, we show for $m = 0 \dots 4$, but we can easily see that the pattern will repeat after $m=6$ (i.e., so that we can take $m$ modulo $6$ to find the rotation angle).}
\label{expfig1}
\end{figure}
\indent 

Quantinuum's TKET compiler \cite{tket} was used to run the circuits---the compiler was set to map the circuit to the physical qubit connectivity and native gateset but not to perform any ``optimization,'' that is circuit rewriting to achieve a functionally equivalent but shallow-depth circuit (compiled circuits for all rounds of experiments are shown in Appendix~\ref{appa}). In this way, we guaranteed that the circuit executed consisted of the appropriate number of repeated Grover iterations, as was necessary to probe the validity of the proposed noise model.

\indent We used minimum mean square error (MMSE) parameter fitting \cite[pp 344-350]{kay1993fundamentals} based on the least-squares method to find the noise-model parameters that best fit the experimental data. That is, each noise model can be expressed as a parameterized function (i.e., measurement outcome probability as a function of number of Grover iterates), and hence, the specific instance of the set of parameters of a given function that minimizes the mean square discrepancy between the experimental values and the values predicted by the noise model was selected as the ``best fit.'' In the fits, we accounted for the statistical uncertainties in the data that arise from the finite number of shots used to estimate the amplitudes.\footnote{It is worth noting that in an earlier analysis (arXiv version 2109.04840v2. [Online]. Available: \url{https://arxiv.org/abs/2109.04840v2}), the estimates were instead treated as point estimates and the final results were similar (particularly expected for the IBM experiments as a large number of data points were averaged over then).}

For each experiment, we fitted three different models: the Gaussian noise model proposed herein; the depolarizing noise model (which is equivalent to the Gaussian noise model with $k_\mu$ set to zero, as shown in Section~\ref{depolsec}); and the Gaussian noise model composed with the amplitude damping channel discussed in Section~\ref{ampdampsec} (where the qubit state that is physically more likely to interact with the environment is specified depending on the particular experiment).

For the first round of experiments, we used four of IBM's five-qubit machines (Athens, Bogota, Rome and Santiago---all of the five-qubit machines online at the time of the experiments), each of which uses transmon superconducting qubits connected in a line \cite{IBM}. For IBM transmon qubits, in the amplitude damping model, the $\ket{0}$ state is mapped to the physical ground state of the system, and the $\ket{1}$ state is mapped to the physical excited state. We ran all of the two-qubit QAE circuits for $m$ applications of $Q$ where $m= 0, \dots, 67$. This range was set on the basis of preliminary studies to discover the reasonable maximum circuit depths (i.e., before noise becomes overwhelming). We ran each circuit for the maximum number of shots allowed at the time, which was 8192.

For the second round of experiments, we ran shots of the circuit $A_{1}$ on three of IBM's five-qubit machines online at the time of the experiments (Lagos, Nairobi, and Perth), each of which also uses superconducting qubits, but these are instead laid out in the shape of a sideways ``H'' \cite{IBM}. We ran 20000 shots of each circuit---the maximum number of shots allowed at the time---and the circuit depths probed were greater than in the first round of IBM experiments, with depths $m= 0, \dots, 80$.

Finally, for the third and fourth rounds of experiments, we used Quantinuum's H1 trapped-ion quantum computer \cite{honeywell}. For Quantinuum's ionic qubits, in the amplitude damping model, the $\ket{0}$ state is mapped to a physical state of the system that is more likely to interact with the environment, and the $\ket{1}$ state is mapped to a physical state of the system that is less likely to interact with the environment.\footnote{This is based on considerations of the H1 device's particular qubit mapping.} Because of the slower gate times and more restricted device availability, we were only able to run a smaller number of shots than when using the IBM devices. However, because of the world-leading (at the time of writing) quantum volume of the Quantinuum machine \cite{honeywellQV}, we were able to confidently run a three-qubit circuit, $A_5$, also shown in Fig.~\ref{Acircs}, which prepares the state given in (\ref{A5defeqn}). 

For the third round of experiments, circuits corresponding to $m = 0, 1, \dots 12$ were run, and for each circuit, 1024 shots were performed. This is denoted ``run 1.'' On a different day, further experiments were run for $m = 0, 1, \dots, 13$, and for each circuit, 1500 shots were performed. This is denoted ``run 2.'' 

For the fourth round of experiments, the circuit depths probed were much greater, corresponding to $m= 0, \dots, 80$, and 1000 shots of each circuit were performed. This is denoted ``run 3.''

The first and second rounds of experiments using IBM hardware were conducted in December 2020, January 2021, and October 2023, respectively, and the third and fourth rounds of experiments using Quantinuum's H1 hardware were conducted in June and July 2021, and August and September 2023, respectively.

\section{Noise-Model Results and Discussion}
\label{res}

\noindent From the MMSE-fitted parameters, we calculate $R^2$ values to evaluate the goodness of fit of the three noise models under consideration. These $R^2$ values are given in Table~\ref{t1}---note that $R^2$ is at most equal to 1 and $R^2 = 1$ means that all of the experimentally observed variance is explained by the proposed model. In addition, while the key quantities for assessing how well the theoretical noise models fit the experimental data are indeed the $R^2$ values, the values of the parameters themselves may also be of general interest, and these are included in Table~\ref{t2} in the appendixes. 

\begin{figure*}[t!] %H if fixed
\centering
\begin{tabular}{ccc}
\includegraphics[width=0.33\textwidth]{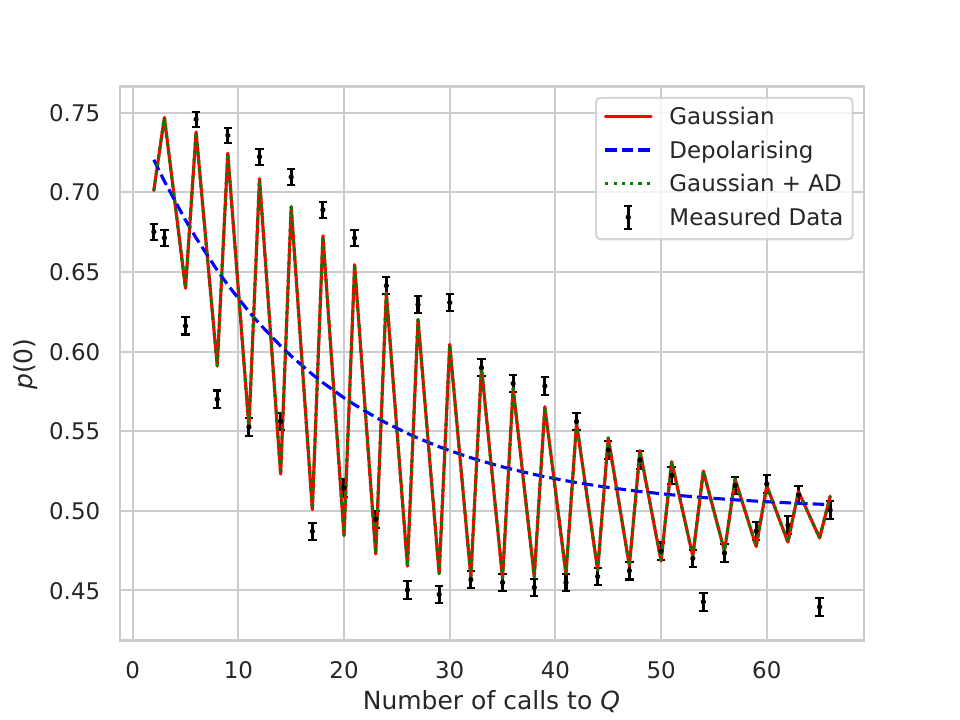} & \includegraphics[width=0.33\textwidth]{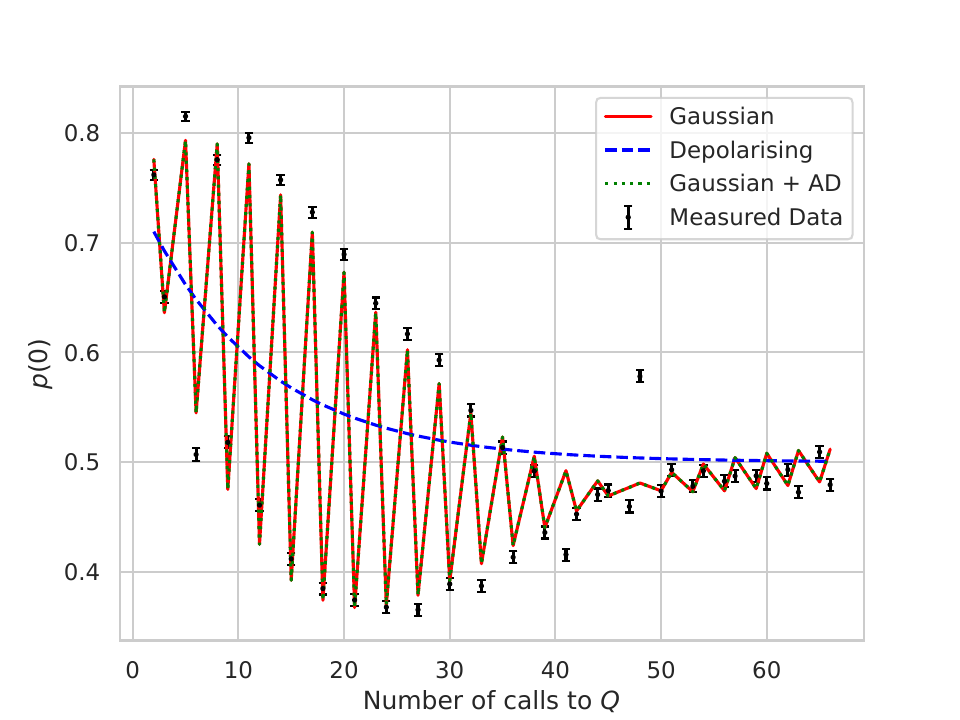} & \includegraphics[width=0.33\textwidth]{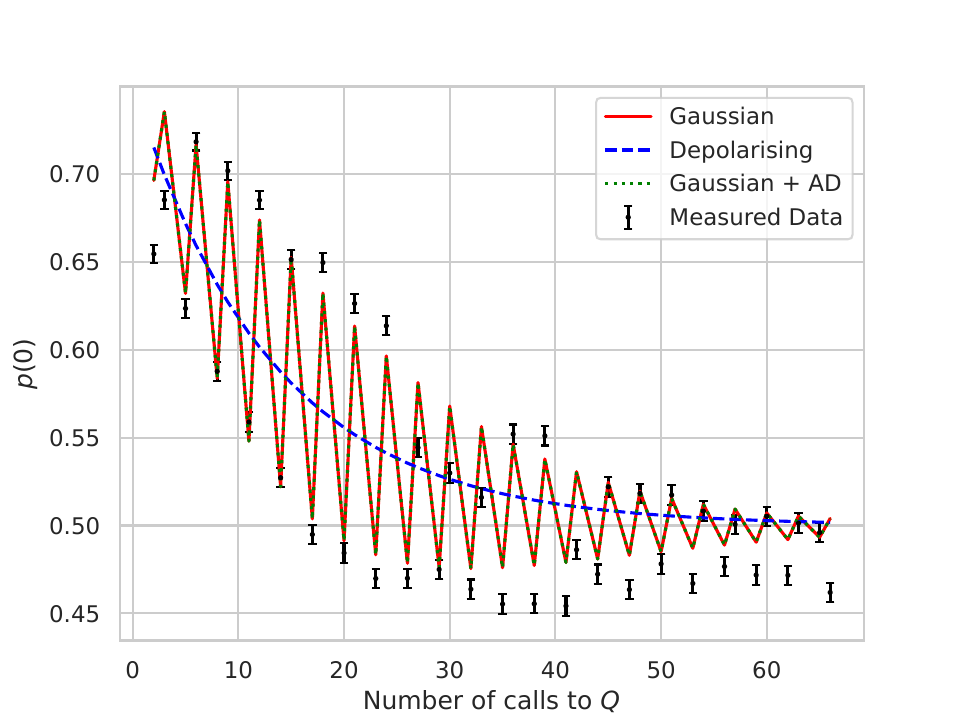} \\
{ } & { } & { } \\
(a) & (b) & (c) \\
{ } & { } & { } \\
\includegraphics[width=0.33\textwidth]{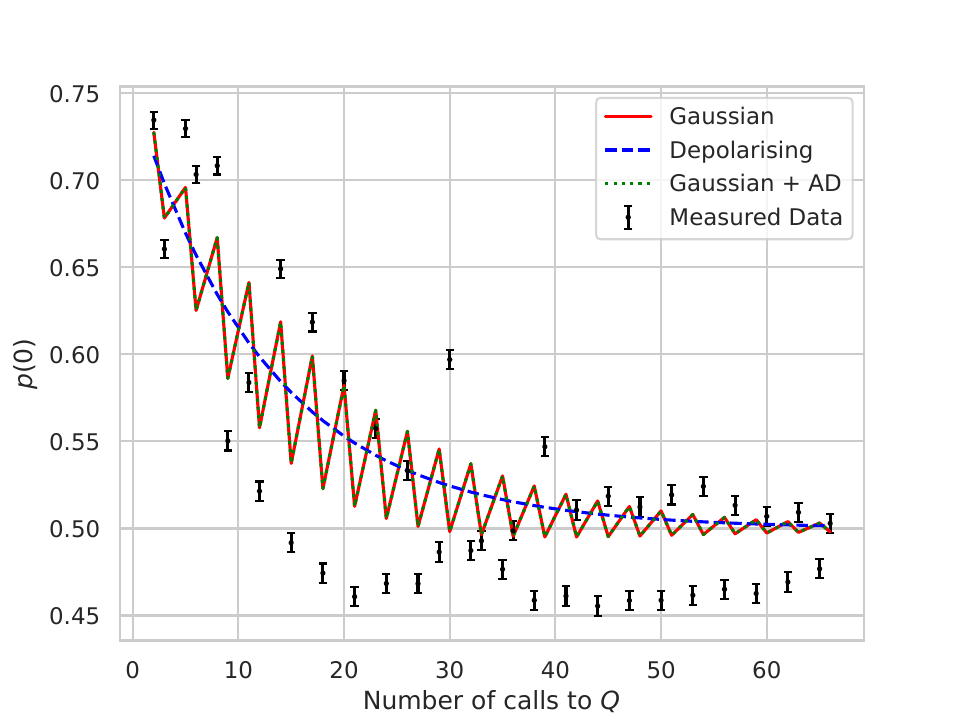} & \includegraphics[width=0.33\textwidth]{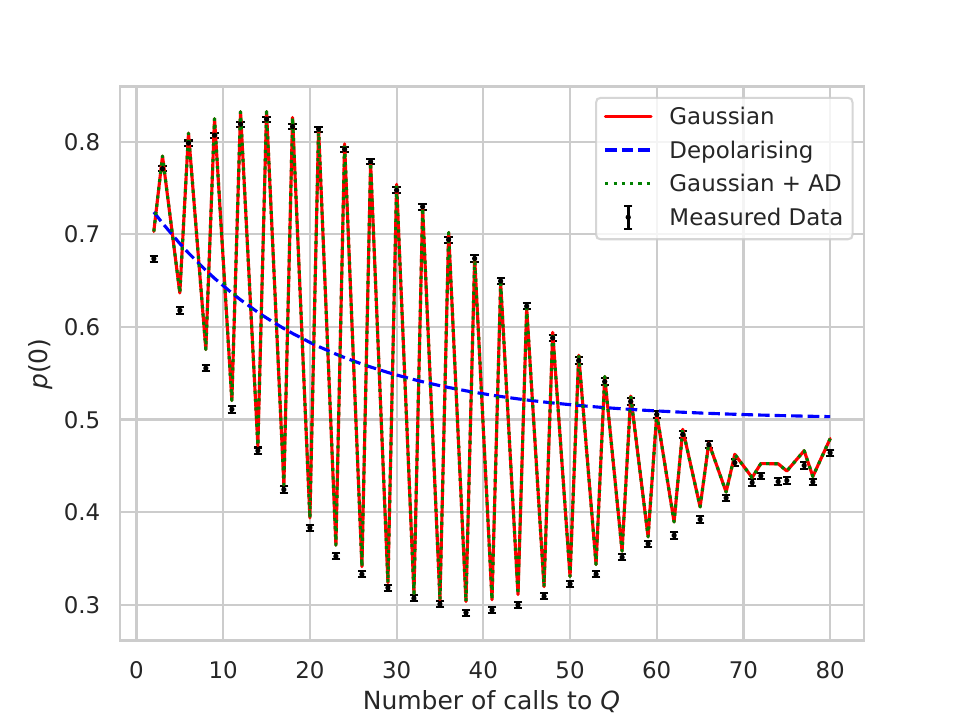} &
\includegraphics[width=0.33\textwidth]{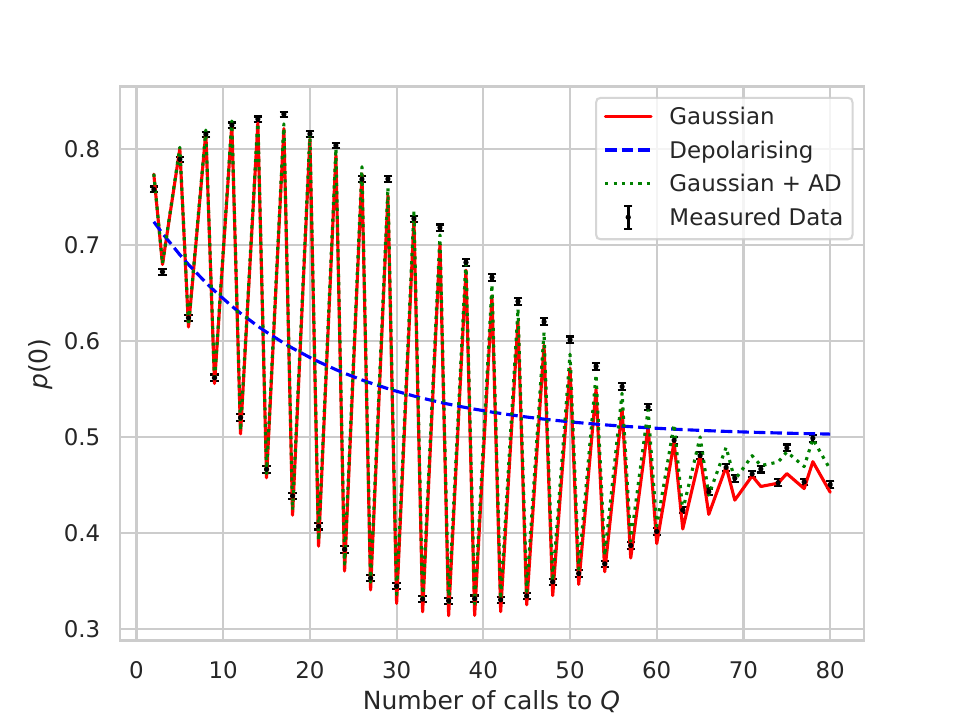} \\
{ } & { } & { } \\
(d) & (e) & (f) \\
{ } & { } & { } \\
 & \includegraphics[width=0.33\textwidth]{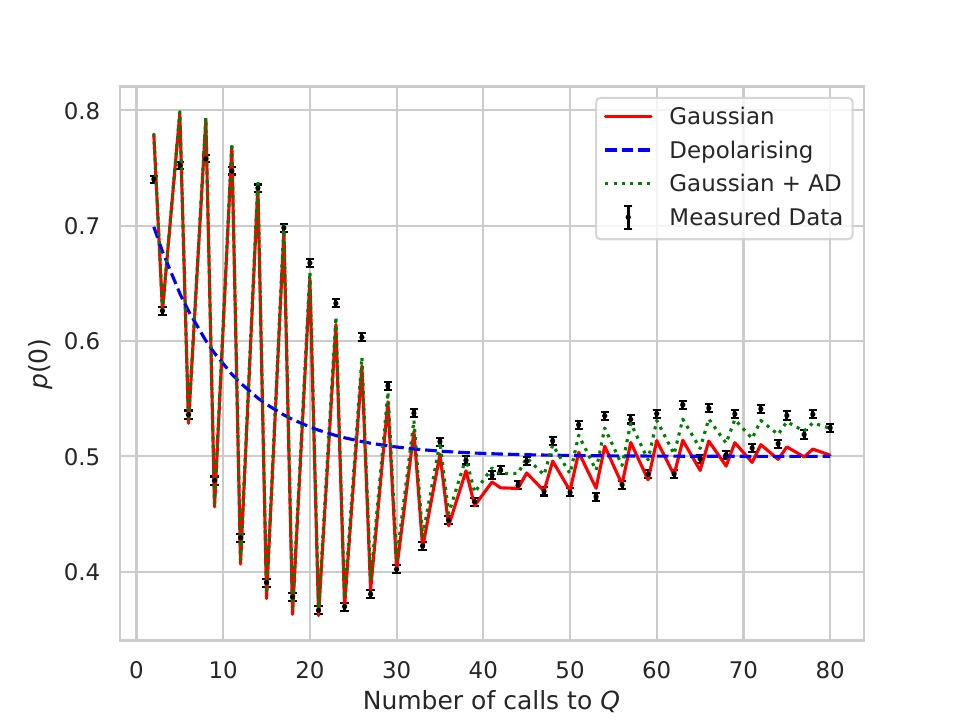} & \\
{ } & { } & { } \\
 & (g) &  \\
{ } & { } & { }
\end{tabular}
\captionsetup{width=0.9\linewidth}
\caption{$A_1$ run on each of the seven IBM machines: results plotted for $2,3,5,6, \dots$ Grover iterations (that is, iterations where in the absence of noise the measurement outcome would be 0 exactly 0.75 of the time). (a) Athens. (b) Bogota. (c) Rome. (d) Santiago. (e) Lagos. (f) Nairobi. (g) Perth.}
\label{resfig2}
\end{figure*}

Fig.~\ref{resfig1} shows plots for $A_1$ run on the various IBM machines, with circuit depths $m=1, 4, 7, \dots$ (i.e., number of Grover iterates such that the noiseless measurement outcome is always equal to 1), and Fig.~\ref{resfig2} shows plots for $A_1$ run on the various machines, with circuit depths $m=0,2,3,5,6,8,9, \dots$ (i.e., number of Grover iterates such that the noiseless measurement outcome is equal to zero with probability 0.75).

\begin{table}[t!]
\center
  \begin{tabularx}{0.95\linewidth}{ c  c  c  c c }
  & & &  \\
 \textbf{Circuit} & \textbf{Machine} & \textbf{Gauss} & \textbf{Dep} & \textbf{AD} \\
  \hline
  \hline
  \multirow{4}{*}{$A_1$} & Athens &   $0.9563$ &  $0.8137$ &   $0.9563$ \\
& Bogota &  $0.9419$  &  $0.5958$ &    $0.9419$ \\
& Rome & $0.9690$ & $0.8772$ &    $0.9690$ \\
& Santiago &  $0.8994$ &  $0.8793$ &    $0.8994$ \\
& Lagos &   $0.9971$   & $0.4663$ & $0.9971$   \\
& Nairobi &   $0.9887$   & $0.4900$ & $0.9943$   \\
& Perth &   $0.9682$   & $0.5784$ & $0.9817$   \\
    \hline
  \multirow{4}{*}{$A_2$} & Athens &  $0.8856$ &  $0.7851$ &  $0.8891$ \\
& Bogota &   $0.8692$  &    $0.7958$ &    $0.8692$ \\
& Rome &  $0.6732$  &  $0.6404$ &    $0.6732$\\
& Santiago &   $0.8975$   & $0.8972$ & $0.8975$   \\
    \hline
      \multirow{4}{*}{$A'_1$} & Athens &  $0.9044$  &   $0.8976$ &    $0.9044$ \\
& Bogota &  $0.8164$  &   $0.7638$ &    $0.8177$ \\
& Rome & $0.6292$   &  $0.6292$ & $0.6292$    \\
& Santiago &    $0.7768$  & $0.7243$ &  $0.7768$    \\
    \hline
          \multirow{4}{*}{$A'_2$} & Athens &  $0.9132$  &   $0.8450$ &    $0.9132$ \\
& Bogota &  $0.7448$   &  $0.6870$ &    $0.7448$ \\
 & Rome & $0.8241$   &   $0.8224$ & $0.8241$   \\
& Santiago &   $0.8026$   &    $0.7939$ &    $0.8026$\\
    \hline
  \multirow{4}{*}{$A_3$} & Athens &  $0.9903$ &   $0.9838$ &    $0.9903$ \\
    & Bogota &  $0.7817$  & $0.6854$ & $0.7998$ \\
    & Rome &  $0.8973$ &   $0.6421$ &    $0.8973$ \\
    & Santiago &   $0.9328$  &   $0.8773$ &    $0.9328$ \\  
    \hline
      \multirow{4}{*}{$A_4$} & Athens &  $0.8890$  & $0.8550$ &    $0.8890$\\
    & Bogota &  $0.8676$  &   $0.7980$ &    $0.8676$ \\
    & Rome &  $0.8040$  &  $0.7301$ &    $0.8040$ \\
    & Santiago &  $0.8648$ &     $0.8217$ &    $0.8648$ \\
    \hline \hline
    \multirow{1}{*}{} &  H1 (Run 1) & $0.9859$  & $0.9854$ &  $0.9928$ \\
    \multirow{1}{*}{$A_5$} &  H1 (Run 2) & $0.9716$ & $0.9667$ & $0.9889$ \\
    \multirow{1}{*}{} &  H1 (Run 3) & $0.9060$ & $0.7920$ & $0.9613$ \\
    \hline
    \hline
  \end{tabularx}
  \captionsetup{width=.9\linewidth}
\caption{$R^2$ values for the various state-preparation circuits running on IBM superconducting quantum computers and Quantinuum's H1 trapped-ion quantum computer. ``Gauss'' stands for ``Gaussian'', ``Dep'' for ``depolarizing'' and ``AD'' for ``Gaussian with amplitude damping''. $A'_1$ and $A'_2$ denote the results from $A_1$ and $A_2$, respectively, when $A_1$ and $A_2$ were run simultaneously.}
  \label{t1}
  
  \end{table}

\begin{figure*}[t!] %H if fixed
\centering
\begin{tabular}{ccc}
\includegraphics[width=0.33\textwidth]{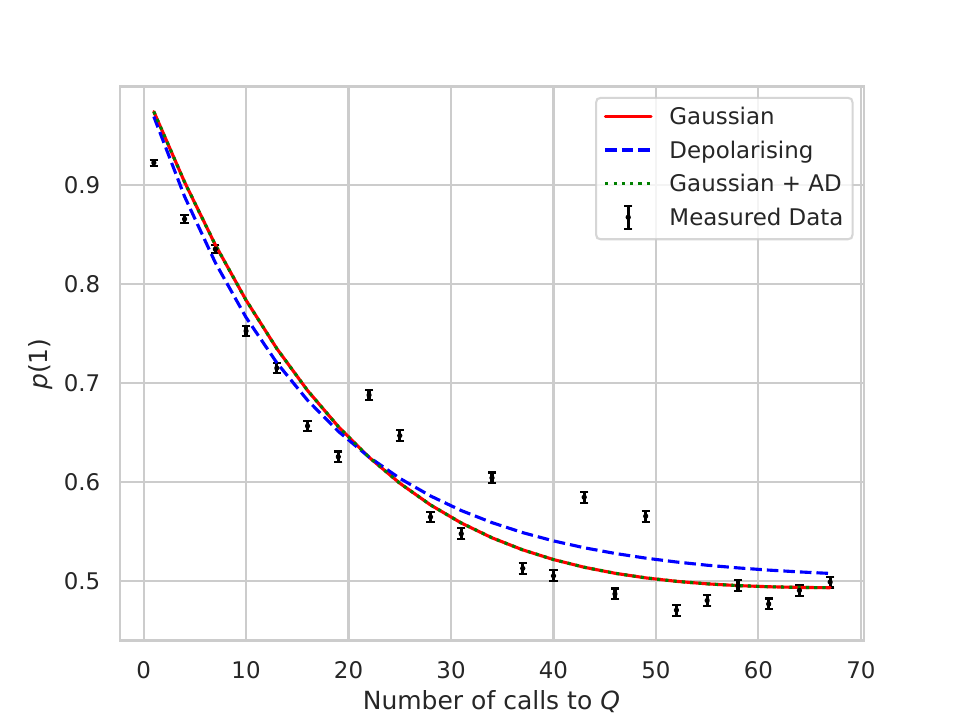} & \includegraphics[width=0.33\textwidth]{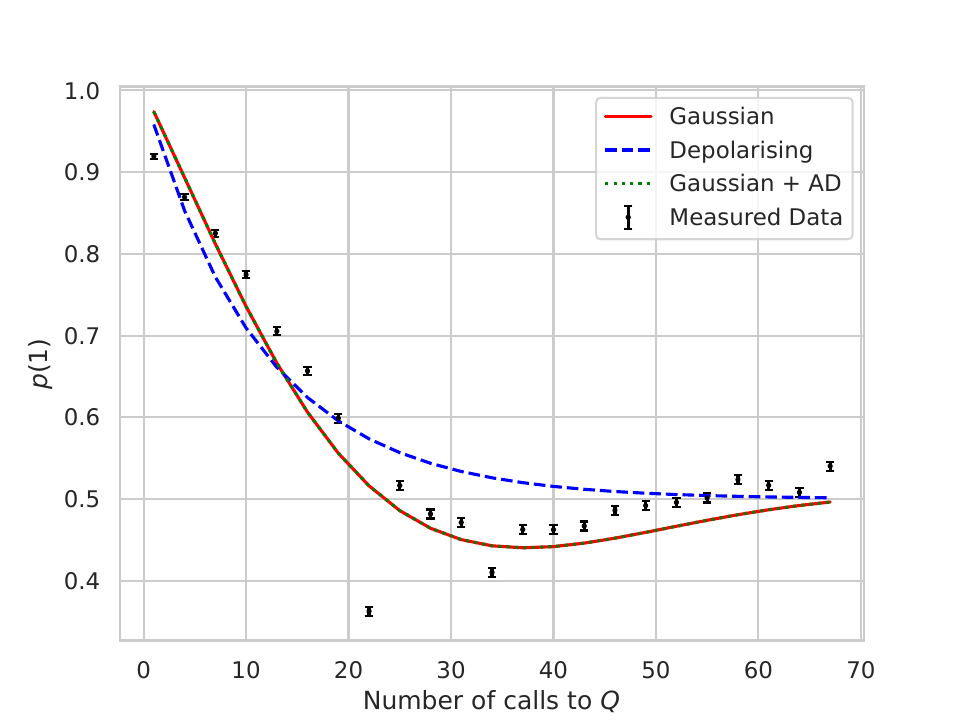} & \includegraphics[width=0.33\textwidth]{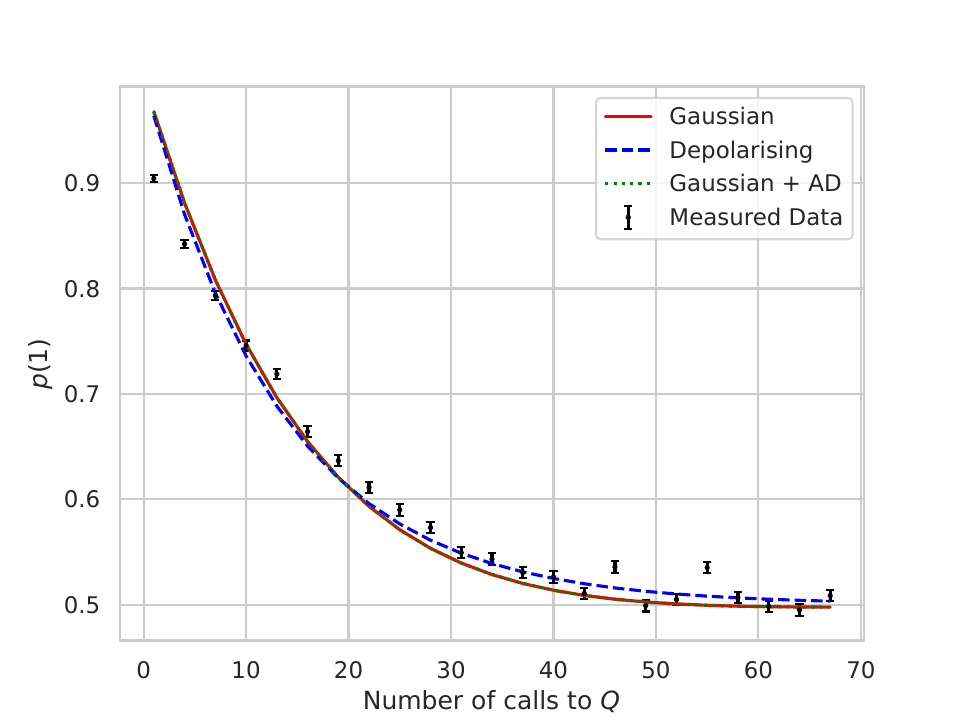} \\
{ } & { } & { } \\
(a) & (b) & (c) \\
{ } & { } & { } \\
\includegraphics[width=0.33\textwidth]{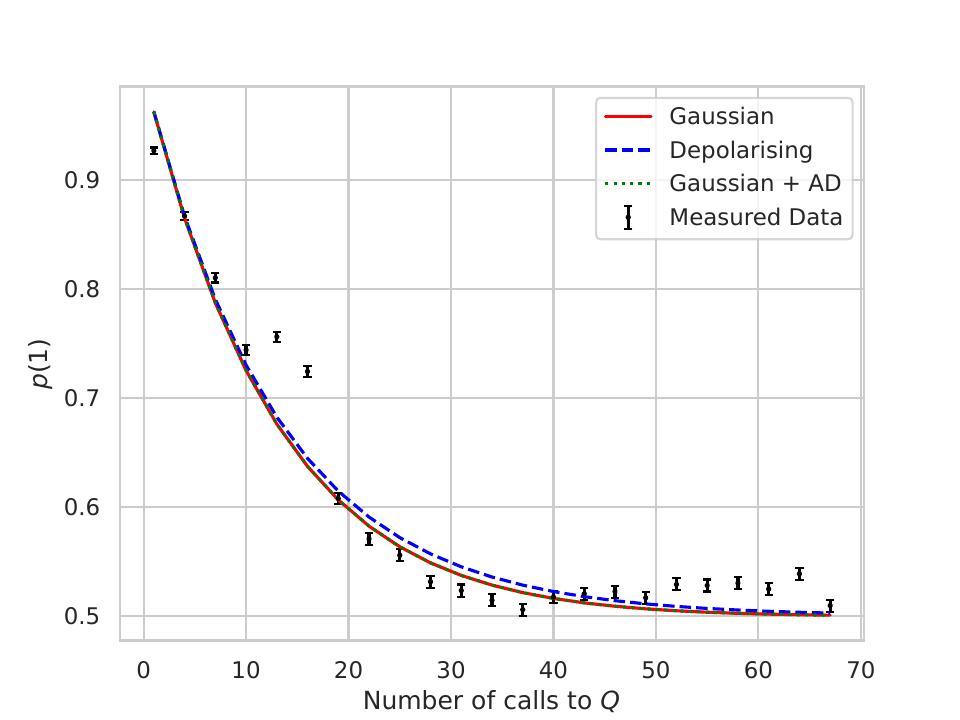} & \includegraphics[width=0.33\textwidth]{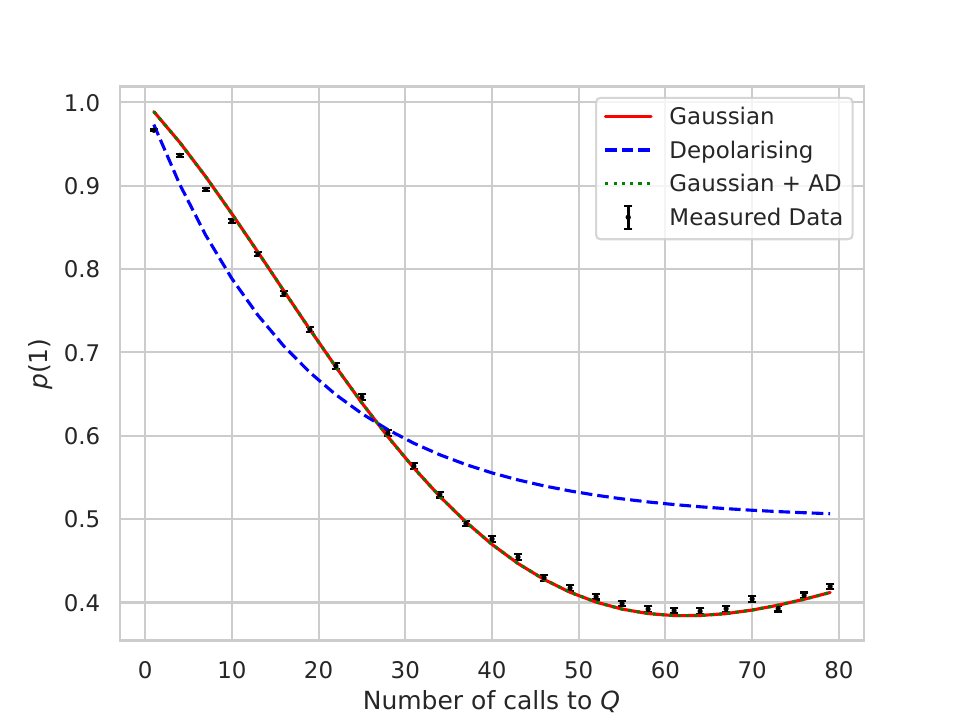} &
\includegraphics[width=0.33\textwidth]{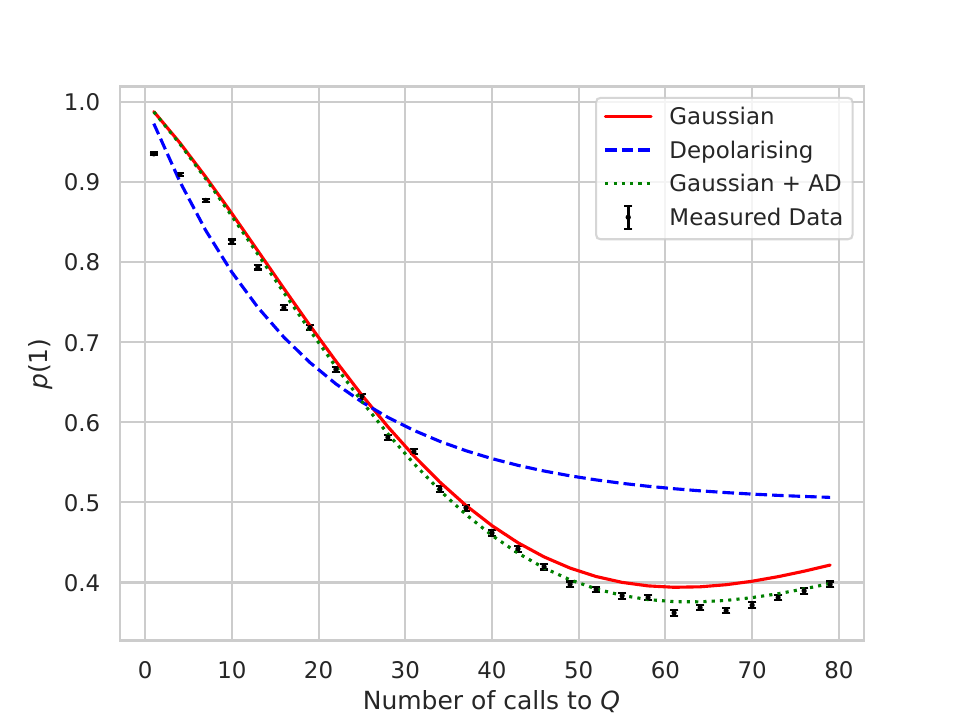} \\
{ } & { } & { } \\
(d) & (e) & (f) \\
{ } & { } & { } \\
 & \includegraphics[width=0.33\textwidth]{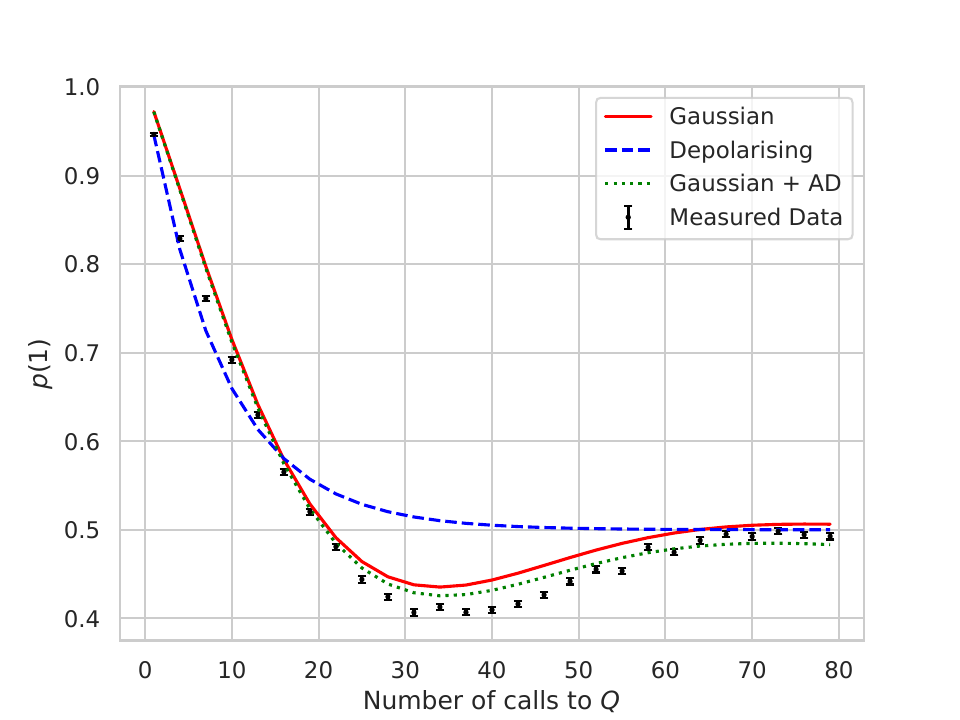} & \\
{ } & { } & { } \\
 & (g) &  \\
{ } & { } & { }
\end{tabular}
\captionsetup{width=0.9\linewidth}
\caption{$A_1$ run on each of the seven IBM machines: results plotted for $1,4,7,\dots$ Grover iterations (that is, iterations where in the absence of noise the measurement outcome would be 1 with certainty). (a) Athens. (b) Bogota. (c) Rome. (d) Santiago. (e) Lagos. (f) Nairobi. (g) Perth.}
\label{resfig1}
\end{figure*}

   \begin{figure*}[t!] %H if fixed
\centering
\begin{tabular}{cc}
\includegraphics[width=0.47\textwidth]{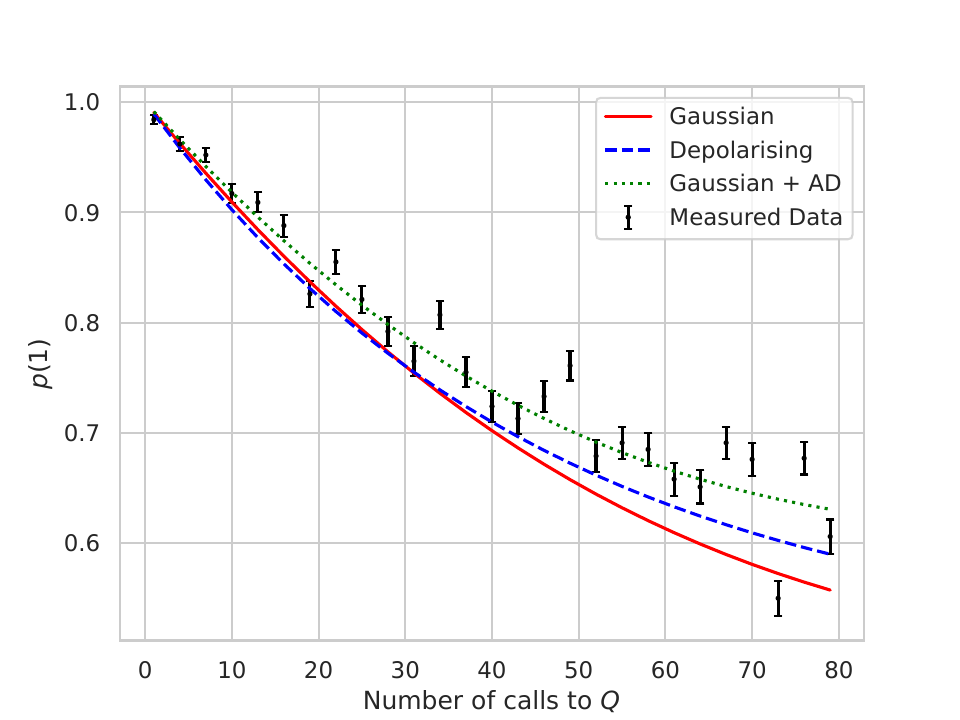} & \includegraphics[width=0.47\textwidth]{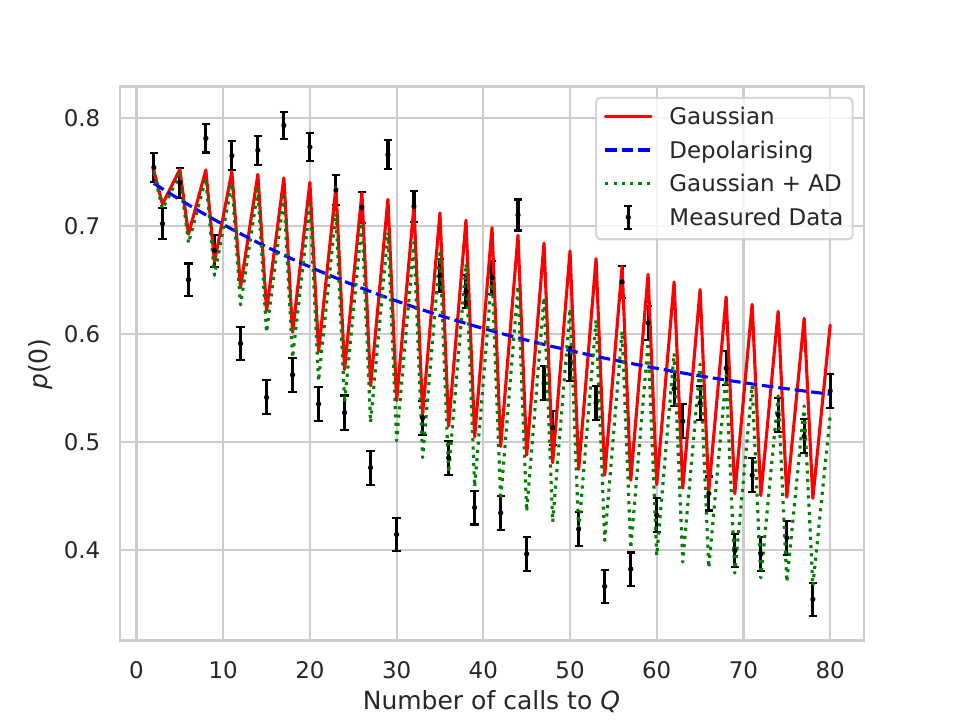} \\
{ } & { } \\
(a) & (b) \\
\end{tabular}
\captionsetup{width=0.9\linewidth}
\caption{$A_5$ run on Quantinuum H1 (Run 3). (a) results plotted for $1,4,7,\dots$ Grover iterations (that is, iterations where in the absence of noise the measurement outcome would be 1 with certainty). (b) results plotted for $2,3,5,6, \dots$ Grover iterations (that is, iterations where in the absence of noise the measurement outcome would be 0 exactly 0.75 of the time).}
\label{honeyfig}
\end{figure*}

   \begin{figure*}[t!] %H if fixed
\centering
\begin{tabular}{cc}
\includegraphics[width=0.47\textwidth]{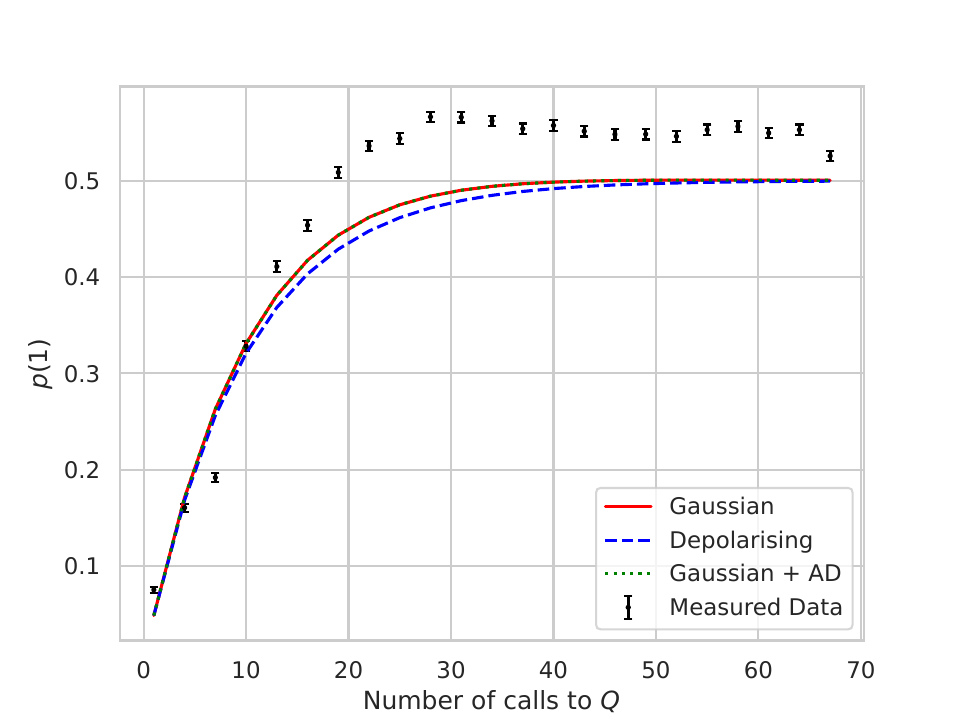} & \includegraphics[width=0.47\textwidth]{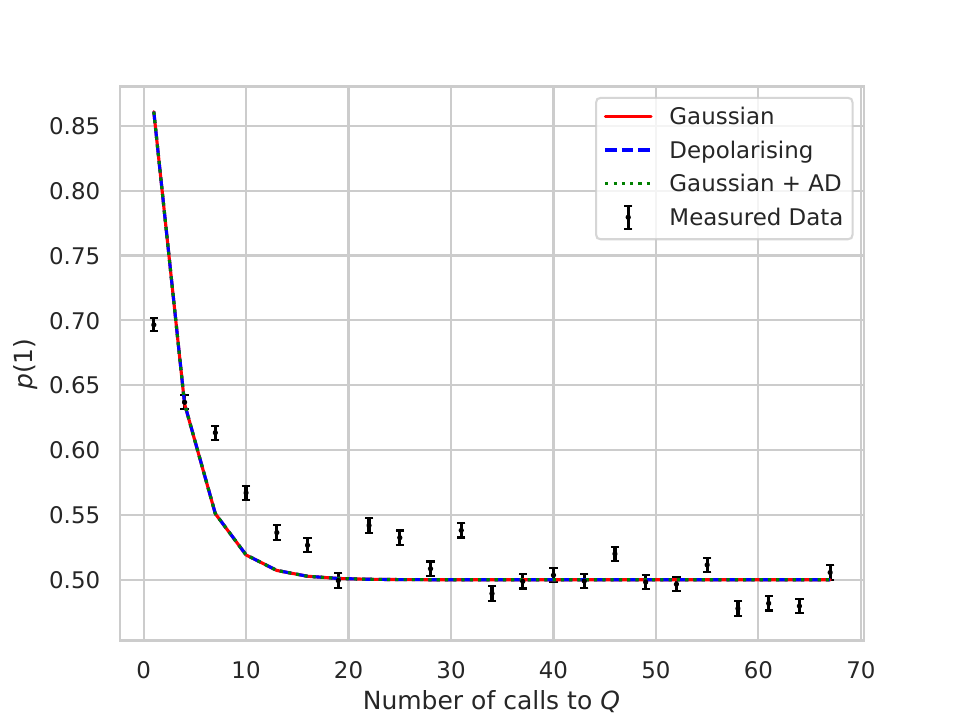} \\
{ } & { } \\
(a)  & (b) \\
\end{tabular}
\captionsetup{width=0.9\linewidth}
\caption{Plots for occasions when $R^2$ indicates that the Gaussian noise model is a poor fit. (a) $A_2$ on Rome ($1,4,7, \dots$ iterations). (b) $A'_1$ on Rome ($1,4,7, \dots$ iterations).}
\label{resfig3}
\end{figure*}

From these results, it is immediately apparent that the proposed Gaussian noise model (with variable mean, and both with and without amplitude damping) explains the experimental data better than the depolarizing noise model (Gaussian noise model with mean fixed at zero)---in almost all cases having (sometimes considerably) higher $R^2$ values. The fact that Gaussian noise model better fits the data can be explained by the fact that the Gaussian noise model with parameterized mean can capture a constant ``bias'' or ``drift,'' where each Grover iterate performs a rotation that is offset from the theoretical rotation angle by some constant amount (plus some zero-mean random noise).  The results for rotation angles $\pi/6$ or $\pi/3$ (as in $A_1$, $A_2$, and $A_5$) clearly illustrate this, as the constant offset is manifested as the probability of measuring zero (or one) being alternately higher and lower (as the number of Grover iterations is incremented) than the simple depolarizing case, as shown in Fig.~\ref{resfig2} (specifically, Fig.~\ref{resfig2} illustrates that noiselessly for all $m$ such that $(m-1) \mod 3 \neq 0$, the probability of measurement outcome being $\ket{0}$ is 0.75). To see an example of this effect, consider the case of a small positive offset to the rotations in Fig.~\ref{expfig1}. In the case of $Q^2$, this offset would move the superposition closer to the direction $-\ket{0}$, and so there would a corresponding increase in the probability of measuring $0$; conversely, in the case of $Q^3$, the offset would move the superposition closer to the direction $-\ket{1}$, and so there would be a corresponding increase in the probability of measuring $1$. Thus, we can see the grounds for the oscillation observed in Fig.~\ref{resfig2}. 

It is worth remarking upon the fact that the presence of this oscillatory behavior demonstrates a deficiency of the depolarizing noise model in certain circumstances. Whereas the depolarizing model aims to capture the overall degradation of the quantum state from a variety of physical effects (as is also our motivation in proposing the Gaussian noise model), it is clearly the case that ``biased'' behavior in the state degradation cannot be captured. That this behavior is evident in both the IBM machines and when the larger three-qubit circuit, $A_5$, is run on Quantinuum's machine---machines with fundamentally different physical manifestations of quantum computation---highlights the necessity of the Gaussian noise model. Most strikingly, for the circuit $A_1$ run on the newer IBM machines (Lagos, Nairobi, and Perth), as well as on Bogota, the Gaussian noise models (with variable mean) have very high $R^2$ values---thus explaining almost all of the experimental data---whereas the equivalent fits for the depolarizing noise model have much lower $R^2$ values. This contrast demonstrates clearly the effect of this constant bias. A similar contrast between the fit results (though the difference is not as large) is seen for the fit to the Run 3 data from the circuit $A_{5}$ run on H1, demonstrated in Fig.~\ref{honeyfig}; this is not surprising given the fact that $A_{1}$ and $A_{5}$ have identical measurement statistics. It should also be noted that, while less amenable to visualization, the same principle explains the better fit of the Gaussian noise model with parameterized mean for $A_3$ and $A_4$.

Comparing the results for the Gaussian noise model and the Gaussian noise model composed with the amplitude damping channel, we see that there is a clear contrast between the results for the older IBM machines and the newer IBM machines, alongside Quantinuum's machine. In the case of the older IBM machines, the fits suggest that the effects of amplitude damping can be generally considered negligible, as in the majority of cases the parameter $k_{AD}$ is found to be negligible, indicating no significant amplitude damping effect observed in the data. By contrast, for the newer IBM machines and the Quantinuum machine, the Gaussian noise model with amplitude damping is a better fit in all cases except Lagos, as the $R^{2}$ results are always greater (significantly so for Quantinuum Run 3). 

The fact that the effects of the constant bias and effects, such as amplitude damping, seem to be more significant for the newer IBM machines and for Quantinuum H1 could plausibly be explained by considering the improved quantum volume of the newer IBM machines as compared to the older generation, and of the world-leading (at the time of writing) quantum volume of the Quantinuum machine. While quantum volume is not the only metric that quantum computing manufactures strive to improve, there has generally been a strong focus on increasing quantum volume, which has been shown to mostly scale as function of the total gate-error magnitude \cite{Baldwin_2022}. Because the combined effect of the dominant sources of noise such as the effect of applying entangling gates will thus have the strongest impact on quantum volume, and will thus be more suppressed on these machines, then subdominant effects will be clearer to distinguish in the data, i.e., they can give a nonnegligible contribution to the overall bias. For the machine with the best quantum volume (H1), then this is clear to see as both the Gaussian noise model and the amplitude damping component considerably improve the fit. By contrast, in the case of the older IBM machines, because the effects of the dominant sources of noise are more significant, then it is likely that the subdominant effects cannot be distinguished from the total noise, i.e., they will be comparatively negligible. Specifically, modeling effects such as amplitude damping in this case would appear to be unnecessary. However, the Gaussian noise model still provides a better fit for these machines, suggesting that the ``grouped effects'' that are reasonably well captured by Gaussian noise are significant in this case.

To further analyze the results, there is merit in discussing the goodness of fit of the proposed Gaussian noise model from both a phenomenological and operational point of view. Regarding the former, the predictions of the noise model clearly do not explain all of the variation observed in the experimental data (that would correspond to $R^2 = 1$ for all experiments), and thus, we can conclude that there are likely underlying physical causes of the variation that are not captured by the model. 

It is worth remarking that the fact that the proposed Gaussian noise model (with and without amplitude damping) does not capture everything is to be expected, as the experiments were explicitly designed to capture raw data from the quantum hardware, and so do not even accommodate the correction of other effects such as SPAM errors, for instance. Indeed, Fig.~\ref{resfig3} shows the two occasions when the $R^2$ value is particularly low, and SPAM error is a plausible explanation for each. In the case of $A_2$ on Rome, we can see that the experimental data has a bias towards measuring the $\ket{1}$ state as the number of Grover iterations grows large (which is the opposite effect to what one would expect from amplitude damping). One possible explanation for this (there could be others) is that there is a readout error, where (in this case) a bit-flip from 0 to 1 is more likely than the converse---which can be thought of as a type of SPAM error. In the case of $A'_1$ on Rome, even for a shallow circuit that does little more than simply prepare and measure a state, we have that $p(1) = 0.7$ rather than $\approx 1$, so clearly---indeed, almost by definition---there is a SPAM error. As discussed previously however, it is in principle possible to compose any number of physically-motivated noise channels (i.e., not just amplitude damping) with the Gaussian noise model in order to be able to capture additional effects such as SPAM, thus making the Gaussian noise model a very versatile and useful tool.

We now turn to the operational benefits of the Gaussian noise model, which are twofold. First, we note that the Gaussian noise model is parameterized in a way that can readily be used in the application---as expounded on in detail in the following sections. This corresponds to our original motivation to model the noise such that it can be handled at the application level. Second, the fact that the proposed noise model does capture the bias is important for calibration, which turns out to be a necessary requirement for running QAE.

\section{Noise-Aware QAE: Calibration}
\label{qaecal}

\noindent For the rest of this article we only discuss the Gaussian noise model without amplitude damping included, for simplicity. Turning our attention to how the Gaussian noise model can be used to improve amplitude estimates, we must firstly deal with the fact that when running QAE, an unknown constant bias (i.e., an unknown $k_{\mu}$) in the result cannot be easily observed. To see this, consider the simple case in which circuits $A\ket{0^n}$ are used to prepare $\ket{\psi} = \cos\theta \ket{\Psi_0} \ket{0} + \sin\theta \ket{\Psi_1} \ket{1}$, in the ideal case, but in practice actually prepare $\cos (\theta + \theta_c) \ket{\Psi_0} \ket{0} + \sin (\theta + \theta_c) \ket{\Psi_1} \ket{1}$ for some nonzero constant $\theta_c$. In this case, averaging a large number of samples will return an estimate converging on $\theta + \theta_c$, i.e., there is no way to distinguish the actual value, $\theta$, from the bias, $\theta_c$. Of course, if $\theta_c$ is nonzero, but known, then it is trivial to subtract this from the averaged value and obtain an unbiased estimate.

A similar principle concerns the effect of $k_{\mu}$ in the QAE circuits. In particular, if one considers the simple scenario in which there is only an unknown bias, i.e., $k_\sigma = 0$, then according to the Gaussian noise model, the prepared states are of the form: $\cos((2m+1)\theta + k_\mu m) \ket{\Psi_0} \ket{0} + \sin ((2m+1)\theta + k_\mu m) \ket{\Psi_1} \ket{1}  =  \cos((2\theta + k_\mu)m+1 ) \ket{\Psi_0} \ket{0}  + \sin ((2\theta + k_\mu)m+1 ) \ket{\Psi_1} \ket{1}$. For large $m$, as are needed for QAE to deliver a quadratic advantage, the total rotation angle is, therefore, dominated by $(2\theta + k_\mu)m$ and so there is again no way to distinguish $\theta$ and $k_\mu$ from samples thereof. This is clearly an issue as such biases (i.e., nonzero values of $k_{\mu}$) are observed to occur in data, as demonstrated in the previous section, and the magnitude of these biases cannot be known a priori for a given circuit; this can, therefore, significantly affect the accuracy of the final estimates output by QAE. However, all is not lost---the Gaussian noise model with parameterized mean is able to capture the bias via the parameter $k_{\mu}$. This feature allows us to propose a general methodology for running QAE while also accounting for this bias, but this requires making use of additional calibration information. 

We now describe a procedure for collecting and analyzing this calibration data. In Section~\ref{naqae} we discuss how this can then be exploited when running QAE. The basic idea is that, given a QAE circuit $A$---whose amplitude is to be estimated by running QAE on a given machine---one can construct a set of calibration circuits $\{A^{'}\}$, where the calibration circuits have similar structures to $A$ but prepare states with known amplitudes. In particular, the calibration circuits should have the same number of two-qubit gates as $A$ (and preferably similar placements of these gates within the circuits), as the assumption is that the error of a final state will be dominated by the effect of entangling gates, and therefore similar-structured circuits containing the same number of two-qubit gates should have similar error profiles to the desired circuit.

The numerical results in Section~\ref{nares} suggest that the calibration technique discussed is effective for the two-qubit circuit used; however, as hardware matures to enable larger-scale noise-aware QAE experiments to be conducted, an important question is how to conduct such calibration in general. The principle that we have used here is that the circuit structure---i.e., number and relative positions of the gates---is the most significant factor determining the values (or range of possible values) of $k_{\mu}$ and $k_{\sigma}$. If this holds for larger circuits, then a suitable strategy will be to take the circuit $A$, assumed to be compiled into a standard gateset of single-qubit gates and CNOTs, and to replace each single-qubit gate with a Hadamard or $S$ gate (say chosen at random). In this way, one will obtain a stabilizer circuit with the same structure, and hence it will be possible to obtain the ground truth of $\theta$ using standard stabilizer simulation techniques and hence to achieve calibration by comparing this to numerical values obtained by running the stabilizer circuits (thus essentially following the same template as set out herein).

A sample of calibration data can be collected by running repeated shots of these calibration circuits on the given hardware---with the circuit submission preferably spaced out across a period of time that is sufficiently long as to maximize the variation in the noise captured by the calibration sample.\footnote{In particular, running the same circuit at different time periods is useful as it allows one to characterize the variation in the noise of the machine purely as a function of time.} By fitting the Gaussian noise model to the calibration data, for example by using MMSE parameter fitting as in the previous section, one can extract distributions of $k_{\mu}$ and $k_{\sigma}$ values that describe the variation in the noise for a given machine. 

In particular, the distribution of $k_{\mu}$ values for the particular circuit run on the particular hardware can then be directly exploited for running QAE at the application level; for example, empirical bounds on the expected maximum size of $k_{\mu}$ can be constructed based on the distribution, and this bound then be propagated into the Gaussian noise model, or instead, the distribution itself can be used directly at the application level. In this article, we discuss both examples.

Additional experimental studies were carried out during November and December 2023 to explicitly demonstrate the methodology and also explore the validity of the calibration protocol. These were based on running two-qubit circuits of the same structure as $A_{1}-A_{4}$, defined previously, using an IBM five-qubit machine online at the time (Perth). First, we note that all of these circuits prepare states of the form
\begin{equation}
A^{'}\ket{00} = \cos(\phi/2)\ket{\Psi_0}\ket{0} + \sin(\phi/2)\ket{\Psi_1}\ket{1}.
\end{equation}
Based on this a ``template'' calibration circuit $A^{'}(\phi)$ is defined, as shown in Fig.~\ref{Calibcirc}. \begin{figure}[t!]
\centering
\ \ \ \ \begin{quantikz}
  \qw & \gate{R_y(\phi)} & \ctrl{1}& \gate{R_y(\psi)} & \qw  \\
  \qw & \qw  & \gate{R_y(\pi)} & \qw & \qw \\
\end{quantikz} \newline
$\ \ \ \ \ A^{'}(\phi)$
\captionsetup{width=.9\linewidth}
\caption{Example template circuit for calibration. Note that the angle $\psi$ has no effect on the state of the final qubit and thus can be set to any value.}\label{Calibcirc}
\end{figure} This template circuit defines the set of calibration circuits $\{A^{'}\}$, which was obtained by randomly sampling 50 different values for the angle $\phi$ in the interval $[0, 4\pi]$ and defining a corresponding calibration circuit based on each sampled value. A total of 20000 shots of each circuit were run, for $m$ applications of $Q$ where $m=0, \dots, 40$.

We ran all 50 circuits 100 different times, i.e., with the intention of running the same circuits at different time periods in order to probe the variation of the noise parameters for a given circuit. However, due to the fair-use queuing system that exists for submitting circuits to IBM hardware, in practice, it was impossible to schedule exactly when a given circuit would run, and occasionally a significant fraction of scheduled jobs for particular circuits that were intended to be run over different periods were found to run all at once, i.e., sequentially. Because the data from these circuits were not particularly useful for characterizing the variation in $k_{\mu}$, we chose to omit all samples from circuits where greater than $50\%$ of jobs were run sequentially from the subsequent analysis. In addition, for some circuits, some scheduled jobs did not complete, and therefore, for certain samples, the total number of repetitions ended up being less than the $100$ requested.\footnote{Unfortunately, it was not possible to submit replacement jobs or resubmit failed jobs because IBM retired these machines shortly after these studies were carried out.} This left $42$ different circuits to analyze with a variable number of repetitions: in total $3724$ independent data points. For these remaining samples, MMSE fitting was again used to find the Gaussian noise-model parameters (i.e., $k_{\mu}$ and $k_{\sigma}$ values as $\theta$ was fixed in the fit to the $\phi/2$ value of the particular calibration circuit) that best fit each data sample. 

\begin{figure}[t!]
\centering
\includegraphics[width = 0.99\linewidth]{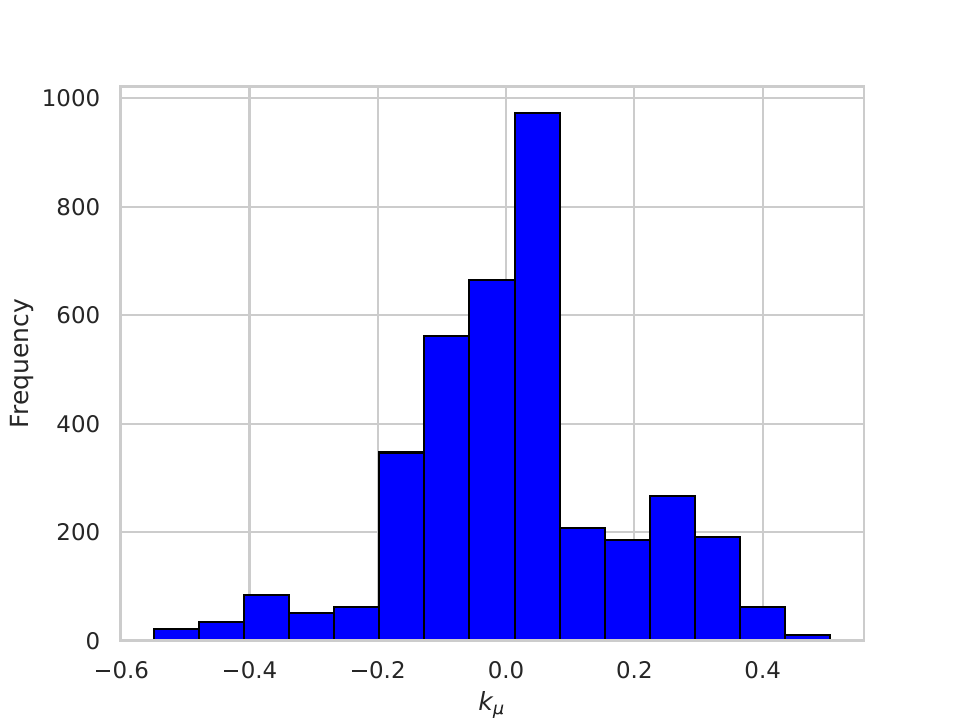} 
\captionsetup{width=.9\linewidth}
	\caption{Distribution of $k_{\mu}$ values across all template circuits and repeated runs for circuits of the form $A_1 - A_4$ run on IBM Perth.}
	\label{kmuhist}
\end{figure}

\begin{figure*} %H if fixed
\centering
\begin{tabular}{c}
\includegraphics[width=0.99\textwidth]{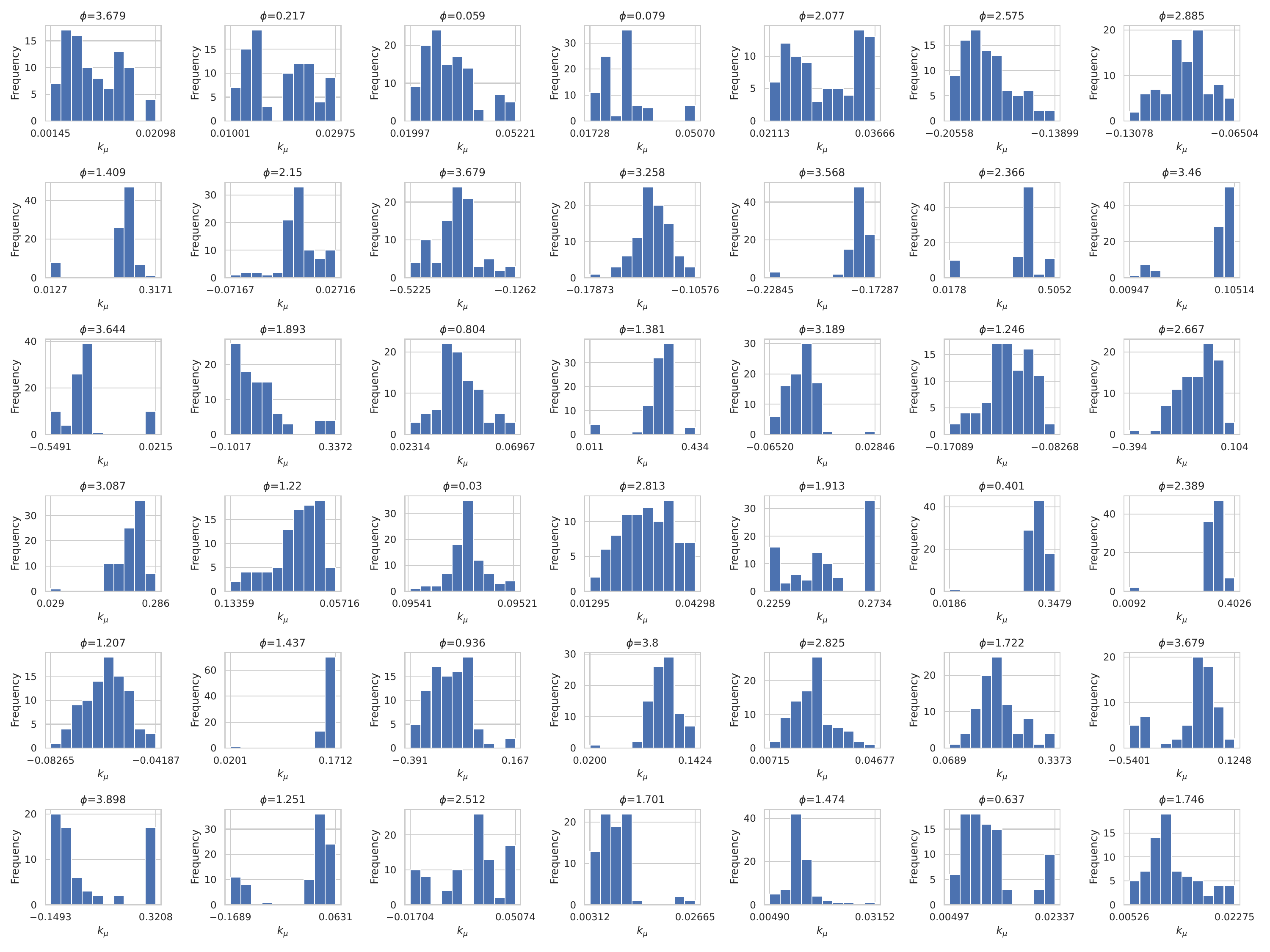} \\
{ } \\
\end{tabular}
\captionsetup{width=0.9\linewidth}
\caption{Distributions of $k_{\mu}$ values for each different template circuit corresponding to a particular $\phi$ value (as a multiple of $\pi$) for circuits of the form $A_1 - A_4$ run on IBM Perth.}\label{kmu_grouped_circuit}
\end{figure*}

\begin{figure*} %H if fixed
\centering
\begin{tabular}{c}
\includegraphics[width=0.99\textwidth]{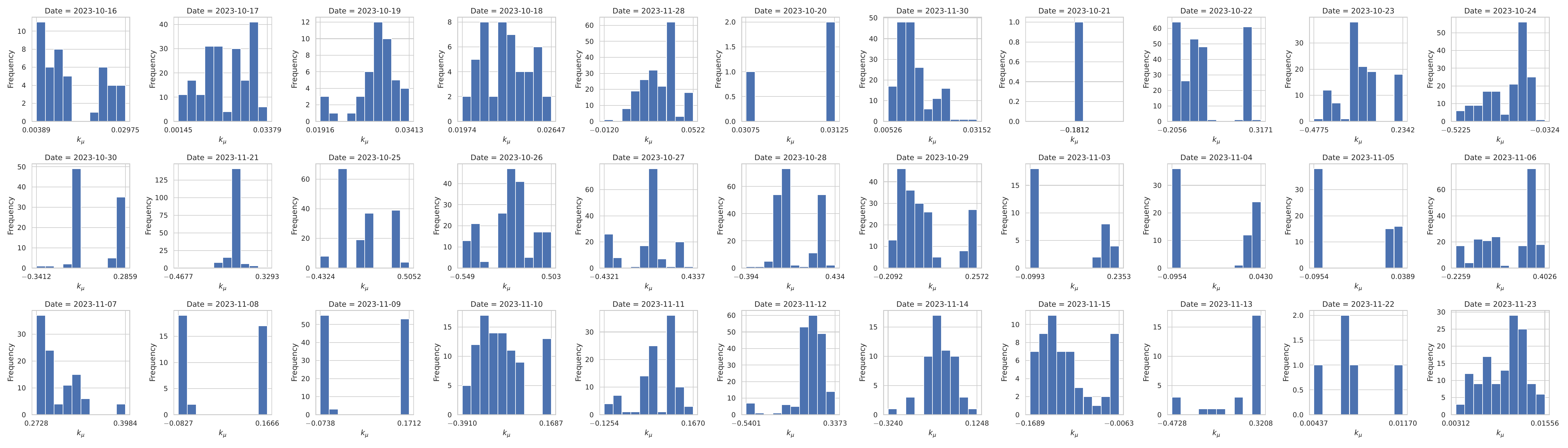} \\
{ } \\
\end{tabular}
\captionsetup{width=0.99\linewidth}
\caption{Distributions of $k_{\mu}$ values for each different 24-h period circuits were run for circuits of the form $A_1 - A_4$ run on IBM Perth.}\label{kmu_grouped_dates}
\end{figure*}

The distribution of $k_{\mu}$ values across all circuits and all repeated runs is given in Fig.~\ref{kmuhist}.  The distribution of $k_{\mu}$ values grouped by the calibration circuit 
and grouped by the particular 24-h period a circuit was run is given in Figs.~\ref{kmu_grouped_circuit} and \ref{kmu_grouped_dates}, respectively.

\
We first focus on the distribution of values across all circuits (i.e., all of the data combined), which can be used to constrain $k_{\mu}$ for the given circuit structure and hardware. From this distribution, the average value of $k_{\mu}$ can be seen to be small (sample mean $\mu_{bias}=0.013$), and the variance is also fairly small (sample standard deviation $\sigma_{bias}=0.169$)---however, clearly, there is a reasonable amount of variation in the parameter. 

In principle, one can use this distribution to determine a bound on the expected maximum absolute size of $k_{\mu}$ for the particular circuit run on the particular hardware, and then, this can be propagated directly into noise-aware QAE. As is usual when handling statistical quantities, there are various ways to use the calibration data for $k_{\mu}$, and the most appropriate one depends on the overall goals of application. For example, if one requires a high level of confidence that any possible bias arising from nonzero $k_{\mu}$ is always accounted for, then a strict bound should be set; however, as described in Section~\ref{naqae}, this typically leads to a large number of additional shots. Assuming that for some application we require a high confidence, then we could directly use the largest absolute magnitude, corresponding to
\begin{equation}
\lvert k^{(strict)}_{\mu} \rvert \leq 0.549\label{eq:strict_bound}
\end{equation} to set a strict upper bound.

Conversely, if a lower confidence is acceptable for the particular application, then a looser bound can be set, with correspondingly fewer additional shots. In particular, we could set a loose upper bound based on the spread around the mean; given that the sample mean and sample standard deviation for $3724$ samples will well estimate the population mean and standard deviation, we could set
\begin{equation}
\lvert k^{(loose)}_{\mu} \rvert \leq  \text{max}(\lvert \mu \pm \sigma \rvert)  =  0.182. \label{eq:loose_bound}   
\end{equation}

Thus, if the priority is to aggressively minimize the total number of samples, then setting a loose bound may be sensible, whereas if the priority is to be highly confident in the precision of a result, then a stricter bound may be sensible. It is worth noting that in Section~\ref{naqae}, we also discuss a different method for constraining $k_{\mu}$ when running QAE at the application level that does not rely on setting a bound and instead uses the distribution directly in a completely data-driven approach.

Moving on to discuss the distributions grouped by the calibration circuit, we see generally that there is a fairly significant variation in the noise profile for a given circuit; this demonstrates that the particular running conditions of the machine have a significant effect on the noise introduced to a given computation. However, it is worth noting that across all circuits the overall variance in these distributions is reasonably small (the largest difference between the largest and smallest value in a sample (range) is $0.665$ and largest standard deviation is $0.181$), and also that the average $k_{\mu}$ values themselves are generally small (the largest absolute average is $0.354$).

Finally, considering the distributions grouped by 24-h running period, we see again that there is a fairly significant variation in the noise profile across each period when running different circuits. This, in contrast to the previous figure, demonstrates that the particular circuits that are run on a given machine also have a significant effect on the noise introduced. Although here again we see that the overall variances are generally small (the largest range is $1.052$ and largest standard deviation is $0.701$), they are larger than those in the previous case. Here again the average $k_{\mu}$ values themselves are generally small (the largest absolute average is $0.301$).

Overall, the grouped distributions demonstrate that the effects of both running conditions and particular circuit structure are important factors for determining the noise, and thus, by varying both across the total calibration sample, we can be confident that our sample is representative of the effects of noise, and can, therefore, be used to characterize $k_{\mu}$.
 
\begin{figure}[t!]
\centering
\includegraphics[width = 0.99\linewidth]{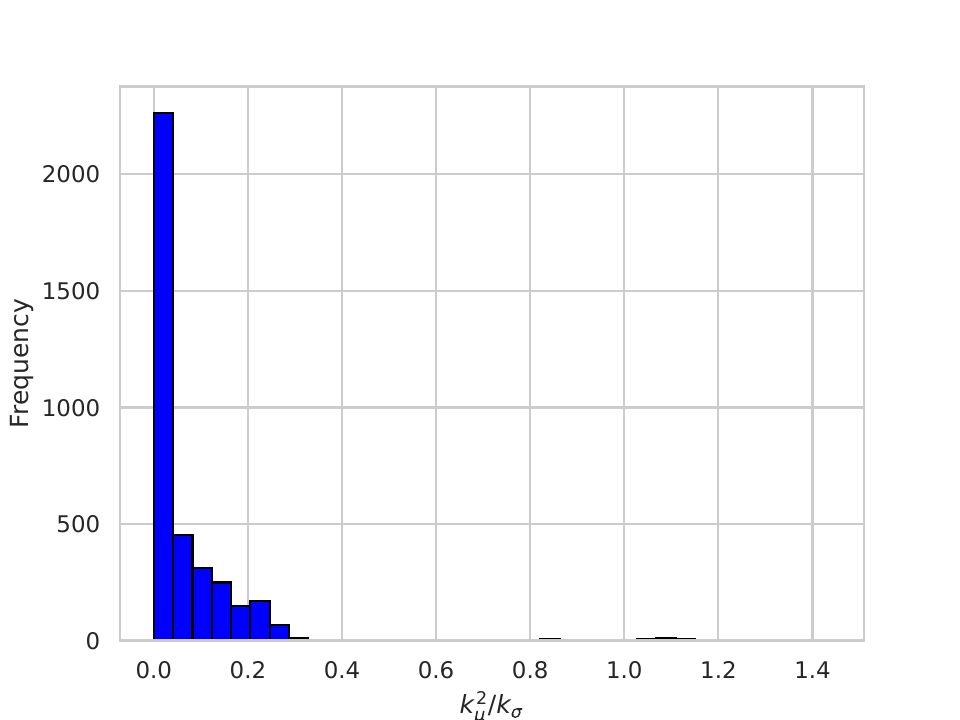} 
\captionsetup{width=.9\linewidth}
	\caption{Distribution of $k_{\mu}^{2}/k_{\sigma}$ values across all template circuits and repeated runs for circuits of the form $A_1 - A_4$ run on IBM Perth.}
	\label{ratiohist}
\end{figure}

In addition, to compare the relative sizes of the parameters $k_{\mu}^{2}$ and $k_{\sigma}$---which is important for considering the total MSE of the estimate when running QAE, as discussed in Section~\ref{naqae}---the distribution of the ratio $k_{\mu}^2/k_{\sigma}$ for all circuits is given in Fig.~\ref{ratiohist}.  Across all samples, this ratio has an average value of $0.069$, which demonstrates that the value of $k_{\mu}^2$ is typically smaller than the value of $k_{\sigma}$, by around an order of magnitude. The full range of ratios varies from approximately $10^{-10}$ to $10^{0}$. However, importantly, only $\sim1\%$ of samples have a ratio that is an order of magnitude or more greater than or equal to the average, and only $\sim0.5\%$ of samples have the ratio being greater than or equal to unity. Based on these findings, it is reasonable to assume, in what follows, that for circuits of the form $A_1-A_4$ run on IBM Perth, that in general $k_{\mu}^2$ and $k_{\sigma}$ differ by around an order magnitude, and that $k_{\mu}^2$ is smaller than $k_{\sigma}$. This fact will be significant in Section~\ref{naqae}.

\section{Noise-Aware QAE: Theory}
\label{naqae}

\noindent We now show how the Gaussian noise model can be used to derive a rule for the necessary factor increase in the number of shots required to counteract the noise at a certain circuit depth, and how it can be used to perform Bayesian inference to estimate $\theta$. To do this, we assume that, even at the maximum circuit depth, the error $\theta_\epsilon$ is a ``small angle,'' that we have collected calibration data for $k_\mu$ for the particular circuit and particular hardware we aim to run based on the methodology outlined in Section~\ref{qaecal}, that this information can be used to construct a strong initial prior probability distribution on $k_\mu$, and that $k_\sigma$ is approximately known from device characterization---here we merely require there to be a weak prior on $k_\sigma$.

Consider that the measurement outcome for a single shot of a circuit with some $m$ Grover iterations is Bernoulli distributed with parameter $\sin^2 \left( (2m+1) \theta \right.$ $\left. + \ \theta_\epsilon \right)$, using the aforementioned small-angle assumption we get
\begin{align}
\sin^2 \left( (2m+1) \theta + \theta_\epsilon \right)  = & \frac{1 -  \cos  \left( 2 \left( (2m+1) \theta + \theta_\epsilon \right) \right)}{2} \nonumber \\
 = & \frac{1}{2} \Big( \! 1 \! - \!  \cos \!  \left( 2 (2m+1) \theta \right) \cos \! \left( 2 \theta_\epsilon\right) \nonumber \\
& \,\,\,\,\,\,  +  \sin \! \left( 2 (2m+1) \theta \right) \sin \! \left( 2 \theta_\epsilon\right) \! \Big) \nonumber \\
 \approx & \frac{1}{2} \Big( \!1 -  \cos  \left( 2 (2m+1) \theta \right) \nonumber \\
 & \,\,\,\,\,\, \,\,\,\,\,\, +  2 \theta_\epsilon \sin  \left( 2 (2m+1) \theta \right) \! \Big) \nonumber \\
 = & \sin^2 \left( (2m+1) \theta  \right) \nonumber \\
 & \,\,\,\,\,\, \,\,\,\,\,\, + \theta_\epsilon  \sin  \left( 2 (2m+1) \theta \right).
\end{align}
Next, we note that, by the Gaussian noise model, $\theta_\epsilon$ is normally distributed, with mean $k_{\mu} m$ (where $k_{\mu}$ has been characterized based on the previous assumption of calibration) and variance $k_ \sigma m$; because $\sin  \left( 2 (2m+1) \theta \right)$ is a constant whose magnitude is at most equal to 1, it therefore follows that $\theta_\epsilon  \sin  \left( 2 (2m+1) \theta \right)$ is also normally distributed
\begin{equation} 
\theta_\epsilon  \sin  \left( 2(2m+1) \theta \right) \sim \mathcal{N} \left( \mu , \sigma^2 \right)
\end{equation}
where $\mu \leq k_{\mu}m$ and $\sigma^2 \leq k_ \sigma m$. 

Next, we let $\alpha_m =  \sin^2 \left( (2m+1) \theta  \right) $ be the amplitude of the qubit after $m$ Grover iterates, and let $N_m$ be the number of shots of this circuit. There are various ways in which the measurement outcomes for all of the different values of $m$ can be combined to infer $\theta$ (and hence $a$), and therefore, to keep things general, here, we simply assume that the objective is to use the $N_m$ shots to provide a point estimate of $\alpha_m$, which we denote $\hat{\alpha}_m$. Taking this approach allows us to determine the increase in the number of shots [i.e., \eqref{algeq1}] to counteract the noise, and the numerical results in Section~\ref{nares} show that it does indeed yield performance improvements. We use the maximum likelihood estimate of $\alpha_m$, which is simply the mean of the $N_m$ samples; using a Gaussian approximation of the binomial distribution, we get that our estimate of the amplitude is normally distributed according to
%
%It is worth noting that this assumption, while simplistic, is likely to be sufficient to give a good guideline for how to increase the number of shots with $m$, certainly in the context of other assumptions that have already been made. 
%
\begin{equation}
\label{aen10}
\hat{\alpha}_m \sim \mathcal{N} \left( \alpha_m + \frac{1}{N_m} \sum_{i=1}^{N_m} \theta^{(i)}_\epsilon  \sin  \left( 2 (2m+1) \theta \right), \tilde{\sigma}^2 \right)
\end{equation}
where $\tilde{\sigma}^2 \leq 1/(4N_m)$ using the fact that a Bernoulli random variable has variance at most one quarter. However, when the mean of a Gaussian is itself normally distributed, it is easy to express the resultant distribution
\begin{equation}
\hat{\alpha}_m \sim \mathcal{N} \left( \alpha_m + \mu, \frac{1}{N_m} \sigma^2 + \tilde{\sigma}^2 \right).
\end{equation} It is worth further remarking that the sum in \eqref{aen10} can be thought of as further loop of summation of the errors, $\epsilon$, associated with each Grover iterate, [as in \eqref{eqn30}] and hence ``compounds'' the use of the CLT, making the assumption of Gaussianity more plausible when the underlying noise model is used in this way.

The MSE of the estimate for $\theta_m = (2m+1)\theta$ is
\begin{equation}\label{mse_eq}
\text{MSE}(\hat{\theta}_m) = \text{Var}(\hat{\theta}_m) + \text{Bias}(\hat{\theta}_m, \theta_m)^{2} = k_{\sigma}m + k_{\mu}^{2}m^{2}.
\end{equation}
If $k_{\mu}$ is exactly known a priori, then it is trivial to adjust the estimator to account for the bias, and hence, the term corresponding to the bias can be set to zero (i.e., in this case $\text{MSE}(\hat{\theta}_m) = k_{\sigma}m$). Letting $N_{shot}$ be the number of shots that would have been selected in the noiseless case (where $k_\sigma = 0$ and $k_{\mu}=0$), 
we can express the factor increase in the number of shots required to obtain a total MSE equal to the noiseless worst case variance for the estimate of $\alpha_m$. In particular, $N_m$ is the number of shots now required when set such that the MSEs for the noisy and noiseless cases are equal 
\begin{equation}
\frac{1}{4 N_{shot}} = \frac{4 k_ \sigma m + 1}{4 N_m} + k_{\mu}^2m^{2}
\end{equation}
when $k_\mu$ is unknown, and
\begin{equation}
\frac{1}{4 N_{shot}} = \frac{4 k_ \sigma m + 1}{4 N_m}
\end{equation}
when $k_\mu$ is known, giving
\begin{align}\label{algeq1}
\begin{cases}
N_m &= \frac{4 k_ \sigma m + 1}{\frac{1}{N_{shot}}-4k_{\mu}^2m^2}, \ \ \ \ \ \ \ \ \   \text{if} \ k_\mu \ \text{is} \ \text{unknown} \\
N_m &= (4 k_ \sigma m + 1)N_{shot}, \ \ \  \text{if} \ k_\mu \ \text{is} \ \text{known}. \\
\end{cases}
\end{align}
From this, the need for the calibration procedure to give some indication of the value of $k_\mu$ is immediately obvious, as the case of ``unknown'' $k_\mu$ has $N_m$ depending on $k_\mu$. Note also that when $k_\mu$ is unknown, the first equation for $N_m$ only has positive solutions up to a particular value of $m$ (where this $m$ value depends on the exact values of $N_{shot}$, $k_{\mu}^2$, and $k_{\sigma}$). This means that after some point, it is not possible to obtain a total MSE equal to the worst case noiseless variance, regardless of how many additional shots are performed. This follows from the fact that at some value of $m$, the component of the total MSE introduced by the bias becomes dominant over the component introduced by the variance due to the former growing quadratically faster as a function of $m$. As the bias component cannot be suppressed by increasing the number of shots, noise-aware QAE is only effective up to some particular value of $m$.

\begin{figure}[t!]
\centering
\includegraphics[width = 0.99\linewidth]{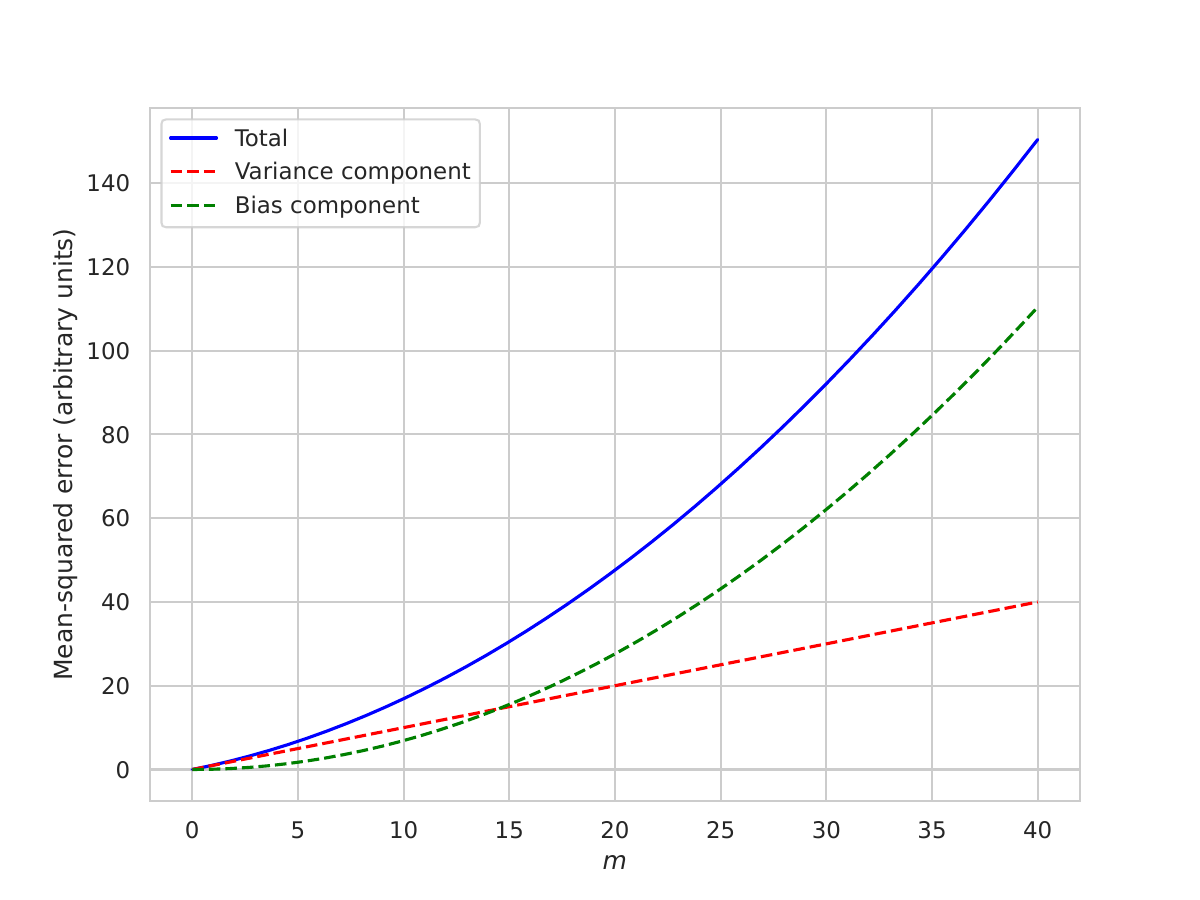} 
\captionsetup{width=.9\linewidth}
	\caption{Average behavior of the MSE as a function of $m$ value for circuits of the form $A_{1}-A_{4}$ run on IBM Perth.}
	\label{mse_plot}
\end{figure}

Turning to the specific case considered in the calibration procedure of Section~\ref{qaecal}, namely circuits of the form $A_{1}-A_{4}$ run on IBM Perth, it was experimentally found that the average ratio of $k_\mu^2 / k_\sigma$ is 0.069. From this, (\ref{mse_eq}) can be used to determine the value of $m$ at which the MSE resulting from the bias begins to dominate by equating the two terms on the right-hand side and solving for $m$, i.e.,
\begin{equation}
    k_\sigma m = k_\mu^2 m^2 \implies m = \frac{k_\sigma }{k_\mu^2} = \left( \frac{k_\mu^2}{k_\sigma} \right)^{-1}
\end{equation}
which, for the average value of $k_\mu^2 / k_\sigma = 0.069$, gives $m=14$. This is confirmed by plotting the average behavior of the MSE alongside the individual components arising from the bias and variance in Fig.~\ref{mse_plot}, from which it can indeed be seen that the contribution to the MSE arising from the bias starts to become significant at approximately $m=14$. This then dictates the approximate regime of validity of our protocol, i.e., where by increasing the number of shots, we can on average expect the total MSE of the estimate to be able to be constrained to be equal to the worst case noiseless case (for circuits of this structure when run on IBM Perth). 

In general, as preempted in Section~\ref{qaecal}, it is unreasonable to suppose that $k_\mu$ and $k_\sigma$ will generally be exactly known a priori, and hence, these must be coestimated alongside $\theta$. To do so, a prior (in the Bayesian sense) on each is required, and for the reasons outlined in the opening of Section~\ref{qaecal}, the prior on $k_{\mu}$ must be relatively strong, to account for the fact that it is hard to distinguish the corresponding offset from the noiseless angle, $(2m+1)\theta$, in circuits of large $m$. To elaborate on this discussion a little, we do expect a strong, but still imperfect prior on $k_{\mu}$ to work in principle (i.e., we do not expect $k_{\mu}$ being exactly known is strictly necessary), as for small values of $m$, the noiseless angle and $k_{\mu}$ can be distinguished, and so a strong (low variance) prior on $k_{\mu}$ with variance further reduced by Bayesian updates from the measurement outcomes of circuits with small $m$ ought to still enable the estimation of $\theta$. 

Conversely, a strong prior is not required for $k_{\sigma}$, as its contribution is completely separable from $\theta$ in \eqref{prob_eqns}, and therefore, the individual uncertainties can be suppressed simultaneously, i.e., if $k_{\mu}$ was equal to zero, then increasing both $m$ and the number of shots will naturally suppress the individual uncertainties in $\theta$ and $k_{\mu}$, respectively.

\begin{figure}[t!]
\centering
\includegraphics[width = 0.99\linewidth]{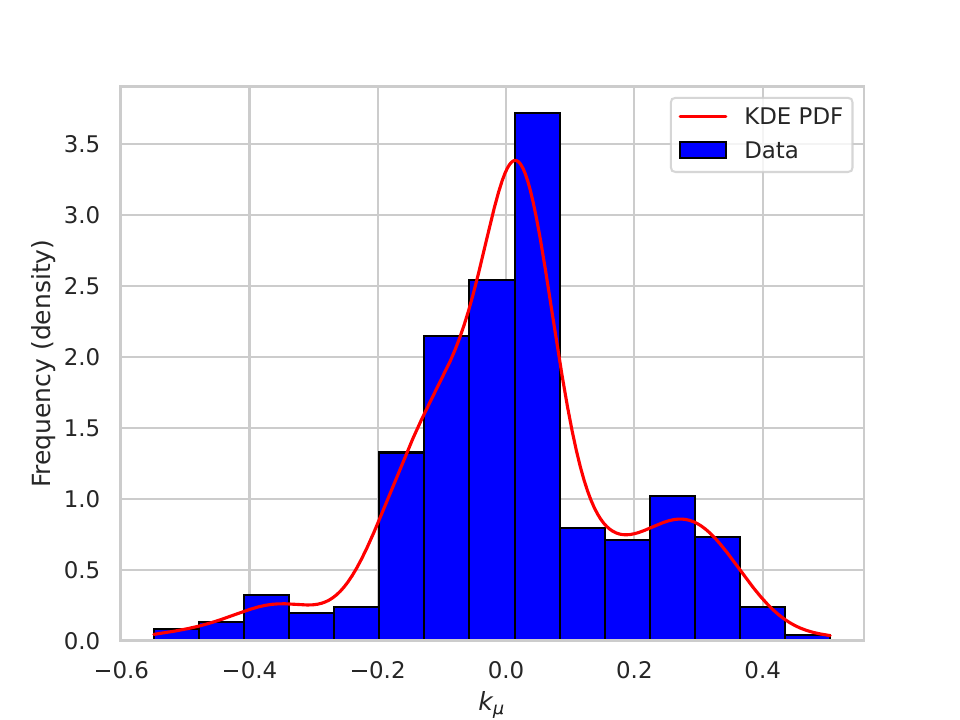} 
\captionsetup{width=.9\linewidth}
	\caption{Densitized distribution of $k_{\mu}$ values for circuits of the form $A_{1}-A_{4}$ run on IBM Perth overlaid with a Gaussian kernel density estimation PDF that approximates the empirical distribution.}
	\label{kmu_kde}
\end{figure}

We now discuss some possible methods to construct the prior for $k_{\mu}$ using the calibration data for $k_{\mu}$, and specifically consider circuits of the form $A_{1}-A_{4}$ run on IBM Perth (i.e., circuits for which we have calibration data). One possible method is to set an uniform prior on $k_{\mu}$ based on the bound set from the calibration data, which, as discussed previously, can either be a strict bound as in \eqref{eq:strict_bound} or a looser bound as in \eqref{eq:loose_bound}, dependent on the particular application. Previously, the use of a uniform prior was chosen for convenience (and in the strict case such that the prior spans the empirically-determined full range of the parameter) rather than because of any phenomenological reasons; hence, we are motivated to consider alternatives too. One such possibility is to take a nonparametric approach and set the prior using a completely data-driven approach, by directly extracting a probability density function (PDF) that approximates the distribution from the data. Because Fig.~\ref{kmuhist} suggests the absence of an evident parametric distribution to describe the data, then employing a nonparametric approach to derive the PDF in this case is reasonable.\footnote{We should note that one would not expect this to necessarily always be the case, and in other cases there may be an appropriate parameterized function that well describes the data.} Utilizing Gaussian kernel density estimation (KDE) \cite{Stanisław18} applied to the densitized distribution, we directly extract a PDF that approximates the PDF in order to use as the prior distribution, as demonstrated in Fig.~\ref{kmu_kde}.  

\begin{algorithm}[!t]
\caption{Noise-aware QAE}\label{alg1}
\begin{algorithmic}[1]
\Require Quantum circuit $A$; prior $p(\theta, k_\sigma, k_{\mu})$; \texttt{Stopping Criterion}
\State Set $m$, $N_{shots}$, $N^{max}_{shots}$
\State From $p(\theta, k_\sigma, k_{\mu})$ obtain point-estimates, $\hat{k}_\mu$ and $\hat{k}_\sigma$ of $k_\mu$ and $k_\sigma$, respectively
\If{\eqref{algeq1} has a valid solution $ \leq N^{max}_{shots}$}
\State Adjust $N_{shots}$ according to \eqref{algeq1} using $\hat{k}_\mu$ and $\hat{k}_\sigma$
\Else{}
\State Set $N_{shots} = N^{max}_{shots}$
\EndIf
\State Prepare and measure $N_{shots}$ of $Q^{m}A\ket{0}$
\State Update $p(\theta, k_\sigma, k_\mu)$ using measurement outcomes
\If{\texttt{Stopping Criterion} is met}
\State Return $\hat{\theta}$
\Else{} Update $m$, $N_{shots}$;
Goto Line 3
\EndIf
\end{algorithmic}
\end{algorithm}

We can bring all this together to give a general framework for embedding noise awareness into any QPE-free QAE algorithm that uses repeated shots of the same circuit to infer the amplitude, as shown in Algorithm~\ref{alg1}. If Lines 2-7 are omitted and $p(\theta, k_\mu, k_\sigma)$ is reinterpreted as simply the marginal probability $p(\theta)$, then we can see that this is a generic framework covering all QPE-free QAE algorithms. As discussed previously, updating the number of shots according to \eqref{algeq1}, as given in Line 4, is only possible up to a particular $m$ value, and therefore, if the algorithm gives no valid solutions for the adjusted number of shots for a particular $m$ value, then $N_{shots}^{max}$ shots are run instead. Clearly, in this case, there are no guarantees to ensure that the total error on the final estimate will be well constrained, and therefore, an actual implementation should warn the user that this is happening, and that, subsequently, the performance may suffer. Likewise, in order that the resources consumed by the user remain bounded, i.e., in order to ensure that the algorithm does not decide to exhaust an unexpectedly large number of shots, if the adjusted number of shots exceeds $N_{shots}^{max}$, then again $N_{shots}^{max}$ shots are run instead (and a similar warning should be given). Both of these conditions are given in Line 3. To avoid the algorithm running infinitely, some (user-specified) stopping criterion should be specified, as checked in Line 10. This could be one (or a combination) of the following: a maximum number of uses of the circuit $A$ has been reached; a maximum value of $m$ has been reached; the desired accuracy has been reached; or some limit on the total runtime has been reached. Note also that normally, but not exclusively, the initial value of $m$ is zero. Also note that all QPE-free QAE algorithms implicitly set a uniform prior for $\theta$ between $0$ and $\pi/2$. It is also the case that this algorithm covers the scenario where either $k_\mu$, $k_\sigma$, or both are known exactly a priori, as this certainty can easily be encoded in the prior, and the rest of the algorithm will play out using the values as constant inputs. It is also worth highlighting that the framework set out in Algorithm~\ref{alg1} captures both adaptive and nonadaptive QAE: in Line 12, the update to $m$ and $N_{shots}$ can either be done by following a preset list, or according to some design based on the current uncertainty of $\theta$. 

\indent It is also important to briefly address the challenge of state preparation for QAE within the context of QMCI, and its connection to noise-aware QAE. The circuit $A$ is constructed from a state-preparation circuit $P$, which encodes a probability distribution for sampling (see \cite{MontanaroMC, herbert2021quantum}). $P$ can represent any arbitrary distribution, and in practical near-term applications, it is crucial to design $P$ to be as shallow as possible. This ensures that the total number of Grover iterations can be maximized, thus improving overall performance. Using a shallow circuit naturally limits the probability distributions that can be prepared in practice. However, noise-aware QAE mitigates this restriction by enabling deeper circuits to be run than would be feasible without. In principle, this allows for sampling from a broader range of distributions. Nonetheless, $P$ should still be designed to be as shallow as possible, as in practice there remains a limit to the effectiveness of noise-aware QAE in terms of the total number of Grover iterations that can be performed, as discussed. Several methods have been proposed for the state preparation of probability distributions (see, e.g., \cite{plesch2011quantum, Sanders_2019, zoufal2019quantum,zhang2022quantum,bausch2022fast,rattew2022preparing,wodecki2024spectral}). One frequently cited approach is the Grover-Rudolph method \cite{GroverRudolph2002}; however, it has been demonstrated that this method offers no quantum advantage when its computational complexity is fully analyzed \cite{herbert2021groverrudolph}. A more recent and promising technique from Rosenkranz et al. \cite{BenedettiRosenkranzStatePrep2024} leverages LCU to prepare multivariate probability distributions. Overall, state preparation for QAE remains an active area of research, and any advancements in this field will greatly enhance the performance of QMCI and other algorithms that make use of QAE as a key subroutine.

\section{Noise-Aware QAE: Results}
\label{nares}

\begin{figure*}[t]
\includegraphics[width=0.95\textwidth]{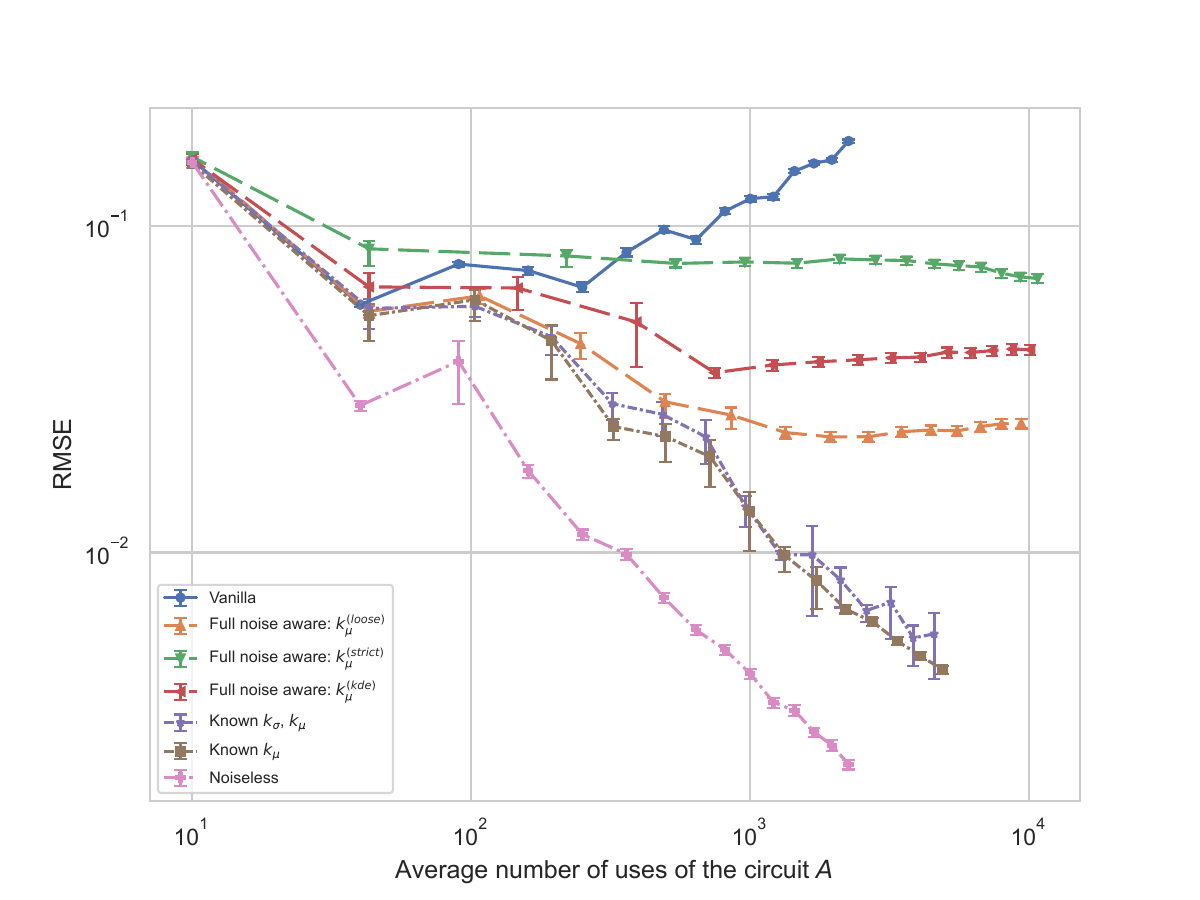}
\caption{QAE for various configurations for the experiment using IBM's Perth five-qubit machine. Bootstrapping---implemented via the bias-corrected and accelerated bootstrap interval \cite{efrondiciccio96}---is used to determine $68\%$ confidence intervals from which the uncertainties on the data points are determined.}
\label{naqaefig}
\end{figure*}

\noindent To assess the performance of noise-aware QAE, we reused the results of the existing experiment running the circuit $A_1$ on IBM's Perth superconducting quantum computer (which is the machine that we have calibration data for), and used the measurement outcome results for each of $\{0,1,2,3,4,5,6,7,8,9,10,11,12,13,14\}$ uses of $A$. This is a ``linearly increasing sequence'' in the terminology of Ref. \cite{Suzuki_2020}. It is important to note that we used this not because it is the best-performing way to run QAE, but because we only considered values up to $m=14$ as this is the average approximate regime of validity for the protocol, as discussed previously, and this way,
we were using the maximum
number of data points in this range (i.e., an m for every integer). However, in reality, we know from Table~\ref{t2} that for this particular dataset, the $k_{\mu}$ and $k_{\sigma}$ values extracted from the entire dataset are the same order of magnitude, i.e., $k_{\mu}$ is larger than average, and therefore, the true regime of validity is below $m=14$; therefore, we may expect the performance to be suboptimal. We also set $N^{max}_{shots} = 5N_{shots}$ as a compromise between  giving enough freedom for the algorithm to be able to run as it should, but also making sure that the algorithm does not run an excessive number of shots for a given $m$. Because the total number of shots run when running the algorithm is nondeterministic, then the total numbers of independent runs of noise-aware QAE that were run varied for the different experimental configurations tested. As Suzuki et al. \cite{Suzuki_2020} only specify that a constant number of shots should be used for each element of the sequence, not a specific value, we chose $10$ as a compromise between final precision and the ability of the algorithm to probe higher $m$ values while still being able to constrain the total error by adjusting the number of shots---we shall see that this is sufficient for the noiseless case. 

We used the experimental data to compare six different configurations\footnote{We note that some of these configurations may not be experimentally possible to realize in reality, i.e., it may not be possible to know the parameter values beforehand. However, we tested them regardless in order to better understand the performance of the algorithm.} alongside simulated data for another configuration, where $N_m = 10$ in all cases:

\begin{enumerate}
\item QAE and no use of the noise model in the parameter estimation, labeled ``Vanilla'';
\item full noise-aware QAE with $p(\theta, k_\sigma, k_\mu)$ set uniformly over $0, \dots, \pi / 2$ for $\theta$, $0.5 k^{(true)}_\sigma, \dots , 1.5 k^{(true)}_\sigma$ for $k_{\sigma}$ (with $k^{(true)}_\sigma$ the ``true'' value from Table~\ref{t2}) and $-\lvert k^{(loose)}_{\mu} \rvert, \dots ,\lvert k^{(loose)}_{\mu} \rvert$ for $k_\mu$ [with $k^{(loose)}$ as given in \eqref{eq:loose_bound}],  labeled ``Full noise aware: $k^{(loose)}_{\mu}$'';
\item full noise-aware QAE with $p(\theta, k_\sigma, k_\mu)$ set uniformly over $0, \dots, \pi / 2$ for $\theta$, $0.5 k^{(true)}_\sigma, \dots , 1.5 k^{(true)}_\sigma$ for $k_{\sigma}$ and $-\lvert k^{(strict)}_{\mu} \rvert, \dots ,\lvert k^{(strict)}_{\mu} \rvert$ for $k_\mu$ [with $k^{(strict)}$ as given in \eqref{eq:strict_bound}],  labeled ``Full noise aware: $k^{(strict)}_{\mu}$'';  
\item full noise-aware QAE with $p(\theta, k_\sigma, k_\mu)$ set uniformly for $\theta$ and $k_\sigma$ as $0, \dots, \pi / 2$ and $0.5 k^{(true)}_\sigma, \dots 1.5 k^{(true)}_\sigma$, respectively, and $k_{\mu}$ set according to the KDE PDF given in Section~\ref{naqae},  labeled ``Full noise aware: $k^{(kde)}_{\mu}$'';
\item noise-aware QAE, but with $k_\mu$ and $k_\sigma$ set to the values extracted from the entire dataset (and so these were not ``learned'' throughout the algorithms run, but set to their ``true'' values $k^{(true)}_\mu$ and $k^{(true)}_\sigma$), labeled ``Known $k_\sigma$, $k_\mu$'';
\item noise-aware QAE with $k_\mu$ set to $k^{(true)}_\mu$ and the marginalized prior $p(\theta, k_\sigma)$ set uniformly over $0, \dots, \pi / 2$ for $\theta$ and $0.5 k^{(true)}_\sigma, \dots , 1.5 k^{(true)}_\sigma$, labeled ``Known $k_\mu$'';
\item simulation of a noiseless quantum computer running QAE, labeled ``Noiseless.''
\end{enumerate}

The results are given in Fig.~\ref{naqaefig}. First, we note that for noise-aware QAE where both parameters were fixed to their ``true'' values, then the algorithm was able to run a deterministic sequence of shots; this number of shots varied as $N_m \! = \! 10, 11, 12, 13, 14, 15, 16, 18, 19, 20, 21, 22, 23, 24, 25$, and we were able to run $800$ independent runs of noise-aware QAE. For all other runs of noise-aware QAE---where one or both of the parameters were learned on the fly---then the numbers of shots run were nondeterministic, and thus, the total number of independent runs available varied for the different configuration tested. 

Focusing on the full noise-aware QAE configurations, we can immediately see that there are clear benefits to using the noise model; without using the noise model (Vanilla), QAE fails to converge at all on the amplitude, whereas the root mean square error (RMSE) can generally (the exception is possibly for ``Full noise aware: $k^{(strict)}_{\mu}$'') be seen to be converging better when the noise model is used to adjust $N_m$ such that more shots are executed when the inclement noise becomes more severe. We see that, as expected, the choice of prior for $k_{\mu}$ has a significant effect on the performance of the algorithm, with the configurations corresponding to setting stronger initial priors (``Full noise aware: $k^{(loose)}_{\mu}$'' and ``Full noise aware: $k^{(kde)}_{\mu}$'') performing much better than the configuration corresponding to the weaker prior (``Full noise aware: $k^{(strict)}_{\mu}$''), which does not appear to really converge at all. Of the three, we see that ``Full noise aware: $k^{(loose)}_{\mu}$'' is the best performing. However, we also see that the convergence of the two performing configurations also begins to flat-line at around $m=5$, and in principle, this is to be expected given the approximate regime of validity of the protocol for this particular dataset, i.e., after $m=5$ there are no valid solutions to \eqref{algeq1}, and therefore, no increase in the number of shots can counteract the bias. This regime of validity, however, does not apply for the configurations with fixed $k_{\mu}$ as the bias is accounted for, and therefore, the number of shots are updated based on the second equation in \eqref{algeq1}; in contrast, we see that in these cases the convergence continues past $m=5$. 

Encouragingly, we see that the performance of the algorithm is fairly agnostic to the assumptions made regarding $k_\sigma$, because even with a relatively high-degree of ignorance about $k_\sigma$ assumed---as is the case for the prior that is set---the performance is comparable between the configurations ``Known $k_\mu$'' and ``Known $k_\sigma$, $k_\mu$'', suggesting that the algorithm is robust to the prior set for $k_\sigma$; we speculate that using published device information (fidelity, quantum volume, etc.) along with coarse information about the circuit, $A$, and making some assumptions regarding the noise, it should be possible to set a prior on $k_\sigma$ which allows noise-aware QAE to work successfully.\footnote{It is likely that the prior itself will not be particularly accurate based on this information alone, but this is likely not an issue when actually running the algorithm, as discussed.} Based on the configurations with fixed parameter values, it also appears to be the case that noise-aware QAE continues to give decent results in practice, even when the small-angle assumption no longer holds: for instance, using the parameter values from Table~\ref{t2}, we get that the RMSE in the angle is about 1.8 rad for $m=10$, which cannot reasonably be called a ``small angle,'' even though there appears to be reasonably decent convergence at this number of Grover iterations.

The final line, ``Noiseless,'' however, shows that even when the parameters are known precisely, there is still an inherent ceiling to the improvement that can be made by employing noise-aware QAE, and also given the variation in the noise parameters, generally one cannot predict a priori how much improvement one might expect for a given problem, particularly given the imperfect method of constraining $k_{\sigma}$ and $k_{\mu}$ from calibration data. However, we still see that there are clear benefits to running noise-aware QAE, and indeed, error mitigation protocols are not expected to improve results to the extent that they match the noiseless case. An important future direction of research is, therefore, to discover effective general ways of setting strong (low-variance) priors on $k_{\mu}$, and indeed to better use the data from circuits with low values of $m$ (i.e., when $\theta$ and $k_{\mu}$ can be better distinguished) to narrow the distribution on $k_{\mu}$ further.

The challenge now is to use the proposed Gaussian noise model, together with ever-improving quantum hardware, to obtain compelling QAE results for problems of real-world interest. Furthermore, the linearly increasing sequence of Suzuki et al. \cite{Suzuki_2020} is known to not be optimal, and is only really used here for illustrative purposes. An important future line of research and development, therefore, will be to embed the noise model into other proposals for QPE-free QAE to obtain noise-aware versions of the state-of-the-art QPE-free QAE algorithms.

\section{Conclusions}
\label{conc}

\noindent QAE is a subroutine that powers many of the most promising potential applications of quantum computing, notably QMCI. In order to run QAE-based algorithms on near-term hardware, it is desirable to find ways to enhance the computational performance in the face of the unavoidable inclement noise. While generic error mitigation techniques may offer some such enhancements, we have identified that the regular structure of QAE circuits offers the opportunity to treat the machine noise as estimation uncertainty, and thus handle the noise at the application level. This approach is attractive as it is tailored to the particular algorithm and offers rigorous performance guarantees even in the face of noisy hardware. 

In this article, we have proposed a simple Gaussian noise model applicable to all near-term QAE algorithms that do not use quantum phase estimation. In addition, we discussed how to extend the model to account for noise effects not well captured by Gaussian noise, such as amplitude damping, by incorporating them into an extended model. We conducted experiments on various IBM superconducting quantum computers and Quantinuum’s H1 trapped-ion quantum computer to validate the model on real hardware. The proposed Gaussian noise model fits the experimental data well, with $R^{2}$ values for the fits to data having a percentage reduction in error (improvement in closeness to unity) for IBM quantum hardware of up to $99\%$ (circuit $A_1$, Lagos), and for Quantinuum quantum hardware of up to $55\%$ (circuit $A_5$, Run 3), compared to fitting a depolarizing noise model. Notably, this is because the Gaussian noise model effectively captures a systematic bias introduced by noise during the application of the Grover iterate in QAE---something generic noise models like depolarizing noise fail to do. Furthermore, we have demonstrated how our noise model can be used to inform the design of, and improve parameter estimation in, QAE, resulting in a noise-aware QAE algorithm. Experimentally, on IBM quantum hardware (Perth), we showed that for a two-qubit circuit with a particular structure, and for a suitable prior distribution on the Gaussian noise parameters, the convergence of amplitude estimation in noise-aware QAE continues to improve up to circuit depths corresponding to approximately five Grover iterates (the expected upper limit of applicability of the methodology given the particular dataset). In contrast, without noise awareness, convergence only improved up to about a single Grover iterate.

An important direction for future research is to combine noise-aware QAE with traditional techniques for error mitigation, and to assess how much further improvement the two complementary approaches give in combination. In addition, another intriguing---although speculative---research avenue would be to explore how noise-aware QAE and QEC might complement each other, both in terms of enhancing QAE performance and reducing the overall resources required for QEC.

\section*{Acknowledgement}

\noindent The authors thank the anonymous reviewers and the editorial team for their careful consideration of the manuscript, Cristina Cirstoiu for helpful conversations leading to the development and refinement of the theoretical model, and Ross Duncan for reviewing. 

\balance


\newpage

\onecolumn

\appendix
\section{Compiled Circuits}
\label{appa}
In this appendix, we give compiled circuits for $A_1 , A_2, A_3, A_4, A_5$.
\begin{figure}[ht!]
\captionsetup[subfloat]{position=bottom,labelformat=empty}
\begin{center}
\subfloat[$A_1$ compiled for IBM (Athens, Bogota, Rome, Santiago)]{\resizebox{1.0\textwidth}{!}{%
\begin{quantikz}
  \qw & \gate{\sqrt{X}} & \gate{R_{z}(3.50)} & \gate{\sqrt{X}} & \gate{R_{z}(1.00)} & \targ{} & \gate{\sqrt{X}} & \gate{R_{z}(2.50)} & \gate{\sqrt{X})} & \gate{R_{z}(1.00)} & \targ{} & \qw & \qw & \qw & \qw & \qw  \\
  \qw & \gate{\sqrt{X}} & \gate{R_{z}(3.33)} & \gate{\sqrt{X}} & \gate{R_{z}(1.00)} & \ctrl{-1} & \qw & \qw & \qw & \qw & \ctrl{-1} & \gate{\sqrt{X}} & \gate{R_{z}(3.13)} & \gate{\sqrt{X}} & \gate{R_{z}(1.00)} & \qw \\
\end{quantikz}}
} \newline
\subfloat[$A_1$ compiled for IBM (Lagos, Nairobi, Perth)]{\resizebox{1.0\textwidth}{!}{%
\begin{quantikz}
  \qw & \gate{\sqrt{X}} & \gate{R_{z}(4.19)} & \gate{\sqrt{X}} & \gate{R_{z}(3.14)} & \ctrl{1} & \gate{R_z(4.71)} & \gate{\sqrt{X}} & \gate{R_z(3.55)} & \gate{\sqrt{X}} & \gate{R_z(3.14)} & \qw \\
  \qw & \gate{R_z(4.71)} & \qw & \qw & \qw & \targ{} & \qw & \qw & \qw & \qw & \qw & \qw \\
\end{quantikz}} 
}\newline
\subfloat[$A_2$ compiled for IBM]{\resizebox{1.0\textwidth}{!}{%
\begin{quantikz}
  \qw & \gate{\sqrt{X}} & \gate{R_{z}(3.50)} & \gate{\sqrt{X}} & \gate{R_{z}(1.00)} & \targ{} & \gate{\sqrt{X}} & \gate{R_{z}(2.50)} & \gate{\sqrt{X})} & \gate{R_{z}(1.00)} & \targ{} & \qw & \qw & \qw & \qw & \qw   \\
  \qw & \gate{\sqrt{X}} & \gate{R_{z}(3.67)} & \gate{\sqrt{X}} & \gate{R_{z}(1.00)} & \ctrl{-1} & \qw & \qw & \qw & \qw & \ctrl{-1} & \gate{\sqrt{X}} & \gate{R_{z}(3.91)} & \gate{\sqrt{X}} & \gate{R_{z}(1.00)} & \qw\\
\end{quantikz} 
}} \newline
\subfloat[$A_3$ compiled for IBM]{\resizebox{1.0\textwidth}{!}{%
\begin{quantikz}
  \qw & \gate{\sqrt{X}} & \gate{R_{z}(3.50)} & \gate{\sqrt{X}} & \gate{R_{z}(1.00)} & \targ{} & \gate{\sqrt{X}} & \gate{R_{z}(2.50)} & \gate{\sqrt{X})} & \gate{R_{z}(1.00)} & \targ{} & \qw & \qw & \qw & \qw & \qw   \\
  \qw & \gate{\sqrt{X}} & \gate{R_{z}(3.32)} & \gate{\sqrt{X}} & \gate{R_{z}(1.00)} & \ctrl{-1} & \qw & \qw & \qw & \qw & \ctrl{-1} & \gate{\sqrt{X}} & \gate{R_{z}(3.13)} & \gate{\sqrt{X}} & \gate{R_{z}(1.00)} & \qw\\
\end{quantikz} }
} \newline
\subfloat[$A_4$ compiled for IBM]{\resizebox{1.0\textwidth}{!}{%
\begin{quantikz}
  \qw & \gate{\sqrt{X}} & \gate{R_{z}(3.50)} & \gate{\sqrt{X}} & \gate{R_{z}(1.00)} & \targ{} & \gate{\sqrt{X}} & \gate{R_{z}(2.50)} & \gate{\sqrt{X})} & \gate{R_{z}(1.00)} & \targ{} & \qw & \qw & \qw & \qw & \qw   \\
  \qw & \gate{\sqrt{X}} & \gate{R_{z}(3.64)} & \gate{\sqrt{X}} & \gate{R_{z}(1.00)} & \ctrl{-1} & \qw & \qw & \qw & \qw & \ctrl{-1} & \gate{\sqrt{X}} & \gate{R_{z}(3.91)} & \gate{\sqrt{X}} & \gate{R_{z}(1.00)} & \qw\\
\end{quantikz} }
} \newline
\subfloat[$A_5$ compiled for Quantinuum H1 (Run 1, Run 2)]{\resizebox{1.0\textwidth}{!}{%
\begin{quantikz}
  \qw & \gate{\text{PhX}(0.25,0.50)} & \gate{R_{z}(1.50)} & \qw & \qw & \gate[wires=3]{ZZ(0.79)} & \qw & \qw & \qw   \\
  \qw & \gate{\text{PhX}(0.25,0.50)} & \gate{R_{z}(1.50)} & \gate[wires=2]{ZZ(0.79)} & \qw & \qw & \qw & \qw & \qw \\
  \qw & \gate{\text{PhX}(0.50,1.50)} & \qw & \qw & \gate{R_{z}(3.50)} & \qw & \gate{\text{PhX}(1.50,0.00)} & \gate{R_{z}(1.50)} & \qw\\
\end{quantikz} }
} \newline
\end{center}
\end{figure}
\newpage
\begin{landscape}
\begin{figure}[ht!]
\captionsetup[subfloat]{position=bottom,labelformat=empty}
\begin{center}
\subfloat[$A_5$ compiled for Quantinuum H1 (Run 3)]{
\resizebox{1.45\textwidth}{!}{%
\begin{quantikz}
  \qw & \gate{\text{PhX}(0.25,0.50)} & \gate{R_{z}(3.91)} & \gate{\text{PhX}(0.50, 0.50)} & \gate{\text{PhX}(1.50, 0.50)} & \gate{R_z(3.91)}  & \qw & \qw & \qw & \qw & \qw & \qw & \qw & \qw & \qw & \qw & \qw & \gate[wires=3]{ZZ\text{Max}} & \gate{R_{z}(1.57)} & \gate{\text{PhX}(0.50, 0.50)} & \gate{R_{z}(3.14)} &  \gate{\text{PhX}(0.50, 0.50)} & \qw   \\
  \qw & \gate{\text{PhX}(0.25,0.50)} & \gate{R_{z}(3.91)} & \gate{\text{PhX}(0.50, 0.50)} & \gate{\text{PhX}(1.50, 0.50)} & \gate{R_z(3.91)} & \qw & \gate[wires=2]{ZZ\text{Max}} & \gate{R_z(1.57)} & \gate{\text{PhX}(0.50, 0.50)} & \gate{R_z(3.14)} & \gate{\text{PhX}(0.50, 0.50)} & \qw & \qw & \qw & \qw & \qw & \qw & \qw & \qw & \qw & \qw & \qw  \\
  \qw & \gate{R_z(3.14)} & \gate{\text{PhX}(0.50,1.00)} & \gate{\text{PhX}(0.50,0.00)} & \gate{\text{PhX}(0.50,0.00)} & \gate{R_z(3.91)} & \gate{\text{PhX}(1.50,0.00)} & \qw & \gate{\text{PhX}(1.50,0.00)} & \gate{R_z(3.91)} & \gate{R_z(3.14)} & \gate{R_z(3.14)} & \gate{\text{PhX}(0.50,1.00)} & \gate{\text{PhX}(0.50,0.00)} & \gate{\text{PhX}(0.50,0.00)} & \gate{R_z(3.91)} & \gate{\text{PhX}(1.50,0.00)} & \qw & \gate{\text{PhX}(1.50,0.00)} & \gate{R_z(3.91)} & \gate{R_z(3.14)} & \qw & \qw  \\
\end{quantikz} 
}}
\end{center}
\end{figure}
\end{landscape}

\begin{landscape}
\begin{center}
A\textsc{ppendix} B. P\textsc{arameter} V\textsc{alues} F\textsc{rom} T\textsc{he} E\textsc{xperiments}
\\
\begin{table*}[h!]
  \begin{tabular}{ c  c  c  c  c c c c c}
  \textbf{Circuit} & \textbf{Machine} & \multicolumn{2}{c}{\textbf{Gaussian}} & 
  \multicolumn{2}{c}{\textbf{Depolarising}} &
  \multicolumn{3}{c}{\textbf{Gaussian with Amplitude Damping}} \\
 & & $k_\mu$ & $k_\sigma$ & $\tilde{p}_{coh}$ & $k_\sigma$ & $k_\mu$  & $k_\sigma$ & $k_{\text{AD}}$  \\
  \hline
  \hline
  \multirow{4}{*}{$A_1$} & Athens &  $-0.0153 \pm 
 0.0002$  & $0.0259 \pm 0.0002$ & $0.9167 \pm 0.0007$ & $0.0315 \pm 0.0002$ & $-0.0153 \pm 0.0002$ & $0.0259 \pm \ 0.0002$ & $0.0000 \pm 0.0000$ \\
& Bogota & $0.0335 \pm 0.0002$  & $0.0255 \pm 0.0002$ & $0.9389 \pm 0.0004$ & $0.0435 \pm 0.0004$ & $0.0255 \pm 0.0002$ & $0.0335 \pm 0.0002$ & $0.0000 \pm 0.0000$ \\
& Rome & $-0.0147 \pm 0.0003$ & $0.0330 \pm 0.0003$ & $0.9277 \pm 0.0006$ & $0.0375 \pm 0.0003$ & $-0.0147 \pm 0.0003$ & $0.0303 \pm 0.0003$ & $0.0000 \pm 0.0000$ \\
& Santiago & $0.0094 \pm 0.0003$ & $ 0.0390 \pm 0.0003 $ & $0.9250 \pm 0.0005$
 & $ 0.0390 \pm 0.0003$ & $0.0094 \pm 0.0003$ & $0.0390 \pm 0.0004$ & $0.0000 \pm 0.0000$\\
 & Lagos & $-0.0214 \pm 0.0000$ & $ 0.0108 \pm 0.0000$ & $0.9465 \pm 0.0002$
 & $ 0.0275 \pm 0.0001$ & $-0.0214 \pm 0.0000$ & $0.0108 \pm 0.0000$ & $0.0000 \pm 0.0000$\\
 & Nairobi & $0.0215 \pm 0.0000$ & $ 0.0115 \pm 0.0000$ & $0.9461 \pm 0.0002$
 & $ 0.0277 \pm 0.0001$ & $0.0021 \pm 0.0000$ & $0.0113 \pm 0.0000$ & $0.0006 \pm 0.0000$\\
 & Perth & $0.0370 \pm 0.0001$ & $ 0.0270 \pm 0.0001$ & $0.8921 \pm 0.0005$
 & $ 0.0571 \pm 0.0003$ & $0.0370 \pm 0.0001$ & $0.0267 \pm 0.0001$ & $0.0006 \pm 0.0000$\\
    \hline
  \multirow{4}{*}{$A_2$} & Athens & $-0.0244 \pm 0.0005$ & $0.0519 \pm 0.0005$ & $0.8791 \pm 0.0012$ & $0.0644 \pm 0.0007$ & $-0.0244 \pm 0.0005$ & $0.0517 \pm 0.0005$ & $0.0003 \pm 0.0000$ \\
& Bogota &  $0.0218 \pm 0.0004$ & $0.0471 \pm 0.0004$ &  $0.9048 \pm 0.0009$ & $0.0500 \pm 0.0005$ & $0.0218 \pm 0.0004$ & $0.0471 \pm 0.0004$ & $0.0000 \pm 0.0000$ \\
& Rome & $0.0173 \pm 0.0005$ & $0.0512 \pm 0.0005$ & $0.9025 \pm 0.0007$ & $0.0513 \pm 0.0004$ & $0.0173 \pm 0.0005$ & $0.0512 \pm 0.0005$ & $0.0000 \pm 0.0000$ \\
& Santiago & $0.0010 \pm 0.0004$  & $0.0412 \pm 0.0003$ & $0.9209 \pm 0.0006$ & $0.0412 \pm 0.0003$ & $0.0010 \pm 0.0004$ & $0.0412 \pm 0.0003$& $0.0000 \pm 0.0000$ \\
    \hline
      \multirow{4}{*}{$A'_1$} & Athens & $-0.0058 \pm 0.0004$ & $0.0445 \pm 0.0004$ & $0.9139 \pm 0.0007$ & $0.0450 \pm 0.0004$ & $-0.0058 \pm 0.0004$& $0.0445 \pm 0.0004$ & $0.0000 \pm 0.0000$ \\
& Bogota & $0.0129 \pm 0.0004$ & $0.0434 \pm 0.0004$ & $0.9130 \pm 0.0007$ & $0.0455 \pm 0.0004$ & $0.0129 \pm 0.0004$ & $0.0433 \pm 0.0004$ & $0.0002 \pm 0.0000$ \\
& Rome &  $-0.0011 \pm 0.0028$ & $0.1630 \pm 0.0028$ & $0.7219 \pm 0.0040$ & $0.1629 \pm 0.0028$ & $-0.0011 \pm 0.0028$ & $0.1630 \pm 0.0028$ & $0.0000 \pm 0.0000$ \\
& Santiago & $0.0164 \pm 0.0003$ & $0.0385 \pm 0.0004$ & $0.9211 \pm 0.0006$ & $0.0411 \pm 0.0003$ & $0.0164 \pm 0.0003$ & $0.0385 \pm 0.0004$ & $0.0000 \pm 0.0000$ \\
    \hline
          \multirow{4}{*}{$A'_2$} & Athens & $-0.0129 \pm 0.0003$ & $0.0331 \pm 0.0002$ & $0.9287 \pm 0.0006$ & $0.0370 \pm 0.0003$ & $-0.0129 \pm 0.0003$ & $0.0331 \pm 0.0002$ & $0.0000 \pm 0.0000$ \\
& Bogota & $0.0241 \pm 0.0007$ & $0.0634 \pm 0.0007$ & $0.8723 \pm 0.0012$ & $0.0683 \pm 0.0007$ & $0.0241 \pm 0.0007$ & $0.0634 \pm 0.0007$ & $0.0000 \pm 0.0000$ \\
 & Rome & $0.0094 \pm 0.0024$ & $0.1438 \pm 0.0022$ & $0.7480 \pm 0.0033$ & $0.1452 \pm 0.0022$ & $0.0094 \pm 0.0024$ & $0.1438 \pm 0.0022$ & $0.0000 \pm 0.0000$\\
& Santiago & $0.0072 \pm 0.0004$  & $0.0424 \pm 0.0004$ &  $0.9198 \pm 0.0092$ & $0.0418 \pm 0.0003$ & $0.0072 \pm 0.0004$ & $0.0424 \pm 0.0004$ & $0.0000 \pm 0.0000$\\
    \hline
  \multirow{4}{*}{$A_3$} & Athens & $-0.0028 \pm 0.0002$ &   $0.0247 \pm 0.0002$& $0.9514 \pm 0.0004$ & $0.0249 \pm 0.0002$ & $-0.0028 \pm 0.0002$ & $0.0247 \pm 0.0002$ & $0.0000 \pm 0.0000$ \\
    & Bogota & $0.0227 \pm 0.0005$  &  $0.0483 \pm 0.0004$ & $0.8932 \pm 0.0011$ & $0.0565 \pm 0.0006$ & $0.0228 \pm 0.0005$ & $0.0480 \pm 0.0004$ & $0.0007 \pm 0.0000$ \\
    & Rome &  $0.0373 \pm 0.0003$  &   $0.0373 \pm 0.0003$& $0.8926 \pm 0.0009$ & $0.0568 \pm 0.0005$ &$0.0353 \pm 0.0003$& $ 0.0373 \pm 0.0003$& $0.0000 \pm 0.0000$ \\
    & Santiago &   $0.0189 \pm 0.0005$ &   $0.0524 \pm 0.0005$& $0.8964 \pm 0.0009$ & $0.0547 \pm 0.0005$ & $0.0189 \pm 0.0005$ & $0.0524 \pm 0.0005$ & $0.0000 \pm 0.0000$ \\  
    \hline
      \multirow{4}{*}{$A_4$} & Athens & $-0.0100 \pm 0.0003$ &   $0.0378 \pm 0.0003$ & $0.9224 \pm 0.0006$ & $0.0404 \pm 0.0003$ & $-0.0100 \pm 0.0003$ & $0.0379 \pm 0.0003$ & $0.0000 \pm 0.0000$\\
    & Bogota & $-0.0132 \pm 0.0004$  &  $0.0418 \pm 0.0004$ & $0.9108 \pm 0.0007$ & $0.0467 \pm 0.0004$ &$-0.0132 \pm 0.0004$& $ 0.04180 \pm 0.0004$& $0.0000 \pm 0.0000$ \\
    & Rome &  $0.0166 \pm 0.0003$  &  $0.0346 \pm 0.0003$ & $0.9264 \pm 0.0006$ & $0.0382 \pm 0.0003$ & $0.0166 \pm 0.0003$& $0.0346 \pm 0.0003$ & $0.0000 \pm 0.0000$ \\
    & Santiago & $-0.0100 \pm 0.0002$   & $0.0282 \pm 0.0002$& $0.9421 \pm 0.0004$ & $0.0298 \pm 0.0002$ & $-0.0100 \pm 0.0002$ & $0.0282 \pm 0.0002$ & $0.0000 \pm 0.0000$ \\
    \hline \hline
\multirow{2}{*}{$A_5$}  &  H1 (run 1) & $-0.0020 \pm 0.0021$ & $0.0545 \pm 0.0020$  & $0.8967 \pm 0.0036$ & $0.0545 \pm 0.0020$ & $-0.0026 \pm 0.0021$ & $0.0522 \pm 0.0021$ & $0.0050 \pm 0.0012$ \\
& H1 (run 2) & $-0.0075 \pm 0.0018$ & $0.0571 \pm 0.0016$  & $0.8915 \pm 0.0029$ & $0.0574 \pm 0.0016$ & $-0.0088 \pm 0.0019$ & $0.0557 \pm 0.0017$ & $0.0072 \pm 0.0009$ \\
& H1 (run 3) & $0.0065 \pm 0.0001$ & $0.0096 \pm 0.0001$  & $0.9784 \pm 0.0004$ & $0.0109 \pm 0.0002$ & $0.0068 \pm 0.0002$ & $0.0087 \pm 0.0001$ & $0.0022 \pm 0.0001$ \\
    \hline
    \hline
  \end{tabular}
  \captionsetup{width=.9\linewidth}
	\caption{\small{Parameter values for the various state-preparation circuits running on IBM superconducting quantum computers and Quantinuum's H1 trapped-ion quantum computer. $A'_1$ and $A'_2$ denote the results from $A_1$ and $A_2$ when $A_1$ and  $A_2$ were run simultaneously, respectively. $\tilde{p}_{coh}$  under the depolarising noise model was calculated from $k_\sigma$ using the established conversion $k_\sigma = - \log_e ( \tilde{p}_{coh} ) / 2$.}}
  \label{t2}
  \end{table*}
 \end{center}
 \end{landscape}

\end{document}